\documentclass[fleqn,usenatbib]{mnras}

\usepackage{newtxtext,newtxmath}

\usepackage[T1]{fontenc}

\DeclareRobustCommand{\VAN}[3]{#2}
\let\VANthebibliography\thebibliography
\def\thebibliography{\DeclareRobustCommand{\VAN}[3]{##3}\VANthebibliography}

\usepackage{graphicx}	
\usepackage{amsmath}	
\usepackage[normalem]{ulem}
\usepackage{rotating}
\usepackage{booktabs}
\usepackage{hyperref}

\title[JWST PAH luminous galaxies]{Polycyclic aromatic hydrocarbon (PAH) luminous galaxies in JWST CEERS data}

\author[Lin et al.]{
Yu-Wei Lin$^{1}$,
Cossas K.-W. Wu$^{1,2}$,
Chih-Teng Ling$^{1,2}$,
Tomotsugu Goto$^{1,2}$, 
Seong Jin Kim$^{2}$,
Ece Kilerci$^{3}$,
\newauthor
Tetsuya Hashimoto$^{4}$,
Po-Ya Wang$^{1}$,
Simon C.-C. Ho$^{5,6,7,8}$,
Tiger Yu-Yang Hsiao$^{9}$,
Bjorn Jasper R. Raquel$^{4,10}$
 \newauthor
, and Yuri Uno$^{4}$
\\
$^{1}$Department of Physics, National Tsing Hua University, 101, Section 2. Kuang-Fu Road, Hsinchu, 30013, Taiwan (R.O.C.)\\
$^{2}$Institute of Astronomy, National Tsing Hua University, 101, Section 2. Kuang-Fu Road, Hsinchu, 30013, Taiwan (R.O.C.)\\
$^{3}$Sabanc{\i} University, Faculty of Engineering and Natural Sciences, 34956, Istanbul, Turkey\\
$^{4}$Department of Physics, National Chung Hsing University, 145, Xingda Road, Taichung, 40227, Taiwan (R.O.C.)\\
$^{5}$Research School of Astronomy and Astrophysics, The Australian National University, Canberra, ACT 2611, Australia\\
$^{6}$Centre for Astrophysics and Supercomputing, Swinburne University of Technology, P.O. Box 218, Hawthorn, VIC 3122, Australia\\
$^{7}$OzGrav: The Australian Research Council Centre of Excellence for Gravitational Wave Discovery, Hawthorn, VIC 3122, Australia\\
$^{8}$ASTRO3D: The Australian Research Council Centre of Excellence for All-sky Astrophysics in 3D, ACT 2611, Australia\\
$^{9}$Department of Physics and Astronomy, The Johns Hopkins University, 3400 N Charles St. Baltimore, MD 21218, USA\\
$^{10}$Department of Earth and Space Sciences, Rizal Technological University, Boni Avenue, Mandaluyong, 1550 Metro Manila, Philippines\\
}

\date{Accepted 2023 December 22. Received 2023 December 22; in original form 2023 May 08}

\pubyear{}

\begin{document}
\label{firstpage}
\pagerange{\pageref{firstpage}--\pageref{lastpage}}
\maketitle

\begin{abstract}
It has been an unanswered question how many dusty galaxies have been undetected from the state-of-the-art observational surveys. \textit{JWST} enables us to detect faint IR galaxies that have prominent polycyclic aromatic hydrocarbon (PAH) features in the mid-IR wavelengths. PAH is a valuable tracer of star formation and dust properties in the mid-infrared wavelength. The \textit{JWST} Cosmic Evolution Early Release Science (CEERS) fields provide us with wavelength coverage from 7.7 to 21 $\mu$m using six photometric bands of the mid-infrared instrument (MIRI). We have identified galaxies dominated by mid-IR emission from PAHs, termed PAH galaxies. From our multi-band photometry catalogue, we selected ten PAH galaxies displaying high flux ratios of $\log(S_{15}/S_{10}) > 0.8$.  The SED fitting analysis indicates that these galaxies are star-forming galaxies with total IR luminosities of $10^{10}$ $\sim$ $10^{11.5}$ $L_{\odot}$ at z $\sim 1$. The morphology of PAH galaxies does not show any clear signatures of major merging or interaction within the MIRI resolution. The majority of them are on the star-formation main sequence at $z \sim 1$. Our result demonstrates that \textit{JWST} can detect PAH emissions from normal star-forming galaxies at $z \sim 1$, in addition to ultra-luminous infrared galaxies (ULIRGs) or luminous infrared galaxies (LIRGs).

\end{abstract}




\section{Introduction}
Polycyclic aromatic hydrocarbons (PAHs) are chemical compounds based on the Benzene structure ($\rm C_{6}H_{6}$, `aromatic ring') composed of hydrogen and carbon atoms \citep{Candian2018}. These molecules are known to absorb UV/optical light and re-emit infrared (IR) photons via vibration modes, which produce broad features at 3.3, 6.2, 7.7, 8.6, 11.3, 12.7 and 17 $\mu$m in the IR spectrum of local galaxies \citep{Tielens2008, Li2020, Kwok2022}.
Due to this property, PAHs serve as good tracers for star formation rate (SFR)
\citep{Rigopoulou1999,Peeters2003,Peeters2006}. Consequently, it is expected that galaxies with high SFRs will exhibit luminous PAH features.
PAH luminous galaxies are expected to have SFRs \citep{Desai2007, Elbaz2011, Takagi2012, Kim2019, Armus2020}, as the overall strength of the PAH emission features is a good tracer of the star-formation activity in galaxies. 
However, at a redshift of $z=1$, the PAH features shift to 6.6, 12.4, 15.4, 17.2, 22.6, 25.4 and 34 $\mu$m, respectively, resulting in a photometric excess in the mid-IR band, which has a high flux density ratio between observation filters such as $S_{15}/S_{10}>10^{0.8}$. The James Webb Space Telescope (\textit{JWST}) is a state-of-the-art infrared (IR) telescope with a 6.5-meter mirror \citep[][]{Gardner2006, Kalirai2018,McElwain_2023,Wright_2023,Gardner_2023}, possessing high sensitivity and spatial resolution. As such, it presents a unique opportunity to select fainter and more distant galaxies displaying mid-IR excess compared to the \textit{AKARI} and \textit{Spitzer} telescopes. In particular, recent studies \citep{Ling2022, Wu2023, Hsiao2023b, Hsiao2023a} have demonstrated the potential for \textit{JWST} observations to provide improved insight into the nature of galaxies.

The majority of PAH galaxies suffer from statistical bias towards more luminous populations, e.g., luminous IR galaxies (LIRGs, $L_{\rm IR} = 10^{11}-10^{12} L_{\odot}$), and ultra-luminous IR galaxies (ULIRGs, $L_{\rm IR} = 10^{12}-10^{13} L_{\odot}$) \citep{Elbaz2005, Elbaz2007}, up to peak of the star-formation rate density (SFRD) \citep{Weedman2008, Shipley2016}. The majority of these galaxies are star-forming galaxies on the main sequence (MS) with a smaller fraction of starbursts (SB) above the MS at different redshifts \citep{Elbaz2007, Elbaz2011, Whitaker2012, Pearson2018, Popesso2023}. However, research on less luminous sources with $L_{\rm IR} < 10^{11} L_{\odot}$, such as normal/faint galaxies at similar/higher redshifts, is limited. In this work, we aim to identify and investigate the properties of such galaxies, including their star formation rates, stellar masses and morphology.

\textit{JWST} is a game changer which has opened up a new era and enables us to observe faint IR galaxies that were beyond the detection limits of previous generation space telescopes such as \textit{AKARI} and \textit{Spitzer}. With \textit{JWST}, we are now able to peer/select normal/faint IR galaxies that have prominent PAHs even at z $\sim 1$. In this work, we show one of the first such analyses using the \textit{JWST} early release observations.

This paper is organised as follows: In Section \ref{sec:data and methods} we present the properties of \textit{JWST} CEERS field data and spectral energy distribution (SED) model. In Section \ref{sec:results}, we show the SED fitting results by \texttt{CIGALE} \citep{Boquien2019}, and we compare the physical properties of PAH luminous galaxies to the literature in Section \ref{sec:discussion}. Finally, our conclusion is given in Section \ref{sec:conclusions}. A flat cosmology with H$_0$ = 70 km s$^{-1}$ Mpc$^{-1}$, $\Omega_m=0.3$ and $\Omega_{\Lambda}= 0.7$  is used. All magnitudes are given in the AB system.

\section{Data and Methods}\label{sec:data and methods}
\subsection{Data}\label{sec:data}
We utilise the \textit{JWST} MIRI 6 bands observations at 7.7, 10.0, 12.8, 15.0, 18.0, and 21.0 $\mu$m (F770W, F1000W, F1280W, F1500W, F1800W, and F2100W) in the Cosmic Evolution Early Release Science (CEERS) fields \citep{Finkelstein2017}. We used the two sub-fields of CEERS (i.e., jw01345-o001 and jw01345-o002) which cover 2 arcmin$^{2}$ each and 4 arcmin$^{2}$ in total. The detection limits in these 6 bands are 0.20, 0.40, 0.79, 1.3, 4.0, and 10 $\mu$Jy from F770W to F2100W, respectively. The details for source extraction are described in \citep{Wu2023}. 
We cross-match \textit{JWST} sources to the Cosmic Assembly Near-infrared Deep Extragalactic Legacy Survey (CANDELS) Extended Groth Strip (EGS) field \citep{Stefanon_2017}. The CANDELS-EGS survey gathered multi-wavelength photometric data ranging from ultraviolet to infrared wavelengths, observed using six different instruments. The UV data ($u*$, $g'$, $r'$, $i$, $z'$) were obtained through Canada–France–Hawaii Telescope (CFHT) MegaCam observations, while the NOAO Extremely Wide-Field Infrared Imager (NEWFIRM) Medium-Band Survey (NMBS) provided additional photometry. The NIR band ($J$, $H$, and $Ks$) data were acquired through CFHT Wide-field InfraRed Camera (WIRCam) observations, while the Hubble Space Telescope (HST) Advanced Camera for Surveys (ACS) and HST Wide Field Camera 3 (WFC3) provided 5 bands of photometry from optical to NIR. Finally, \textit{Spitzer} IRAC observations provided 4 bands of photometry from 3 to 8 $\mu$m,
detail of the catalogue is described in \cite{Stefanon_2017}. In Table \ref{tab:bands}, we present the main properties of the 21 bands data set which we used to perform the SED fitting. 

\begin{figure}
	\includegraphics[width=\columnwidth]{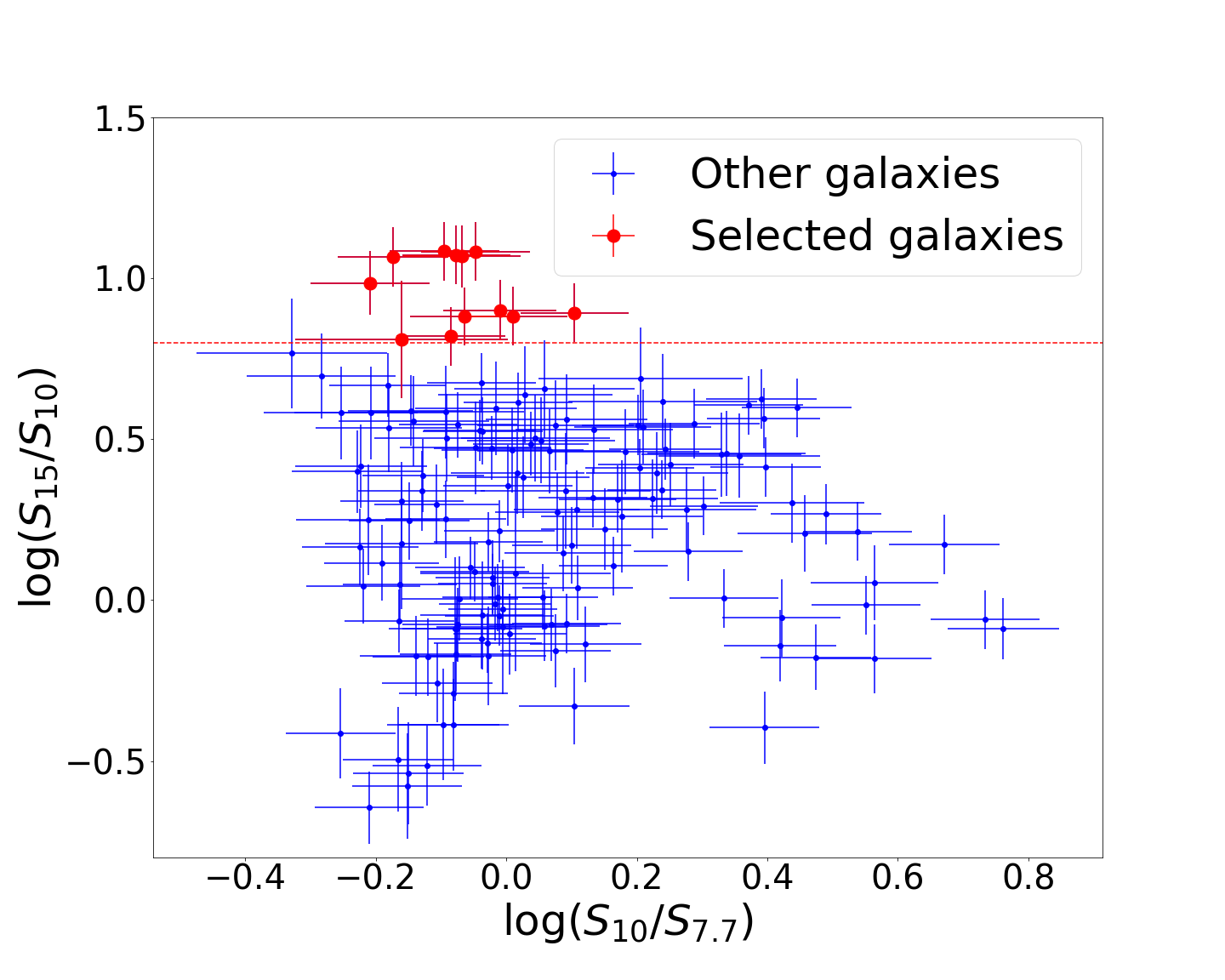}
    \caption{A colour-colour plot of three-bands-detected (7.7, 10, and 15 $\mu$m) source with 15-to-10 $\mu$m and 10-to-7.7 $\mu$m flux ratios. The red dashed line indicates the criteria cut to select the PAH sources from \citet{Takagi2010}. The red points indicate the PAH-selected galaxies.}
    \label{fig:ccplot}
\end{figure}

\subsection{Sample selection}
In Fig. \ref{fig:ccplot}, we show the colour-colour plot between $S_{15}/S_{10}$ and $S_{10}/S_{7.7}$. We select sources based on detection in 3 MIRI bands F770W, F1000W, and F1500W. There are 155 sources in these three bands. We utilise the method from \citet{Takagi2010} to select 15 $\mu$m flux density excess sources that are likely PAH galaxies. The selection criterion for the sample is the ratio of flux density between 15-to-10 $\mu$m greater than 0.8 (dex), or $\log(S_{15}/S_{10}) > 0.8$ from \citep{Takagi2010}, indicated by the red dashed line in Fig. \ref{fig:ccplot}. In total, we identify 12 galaxies that meet the selection criteria.
 
\begin{table*}
    \centering
    \caption{Summary of 21 bands for \texttt{CIGALE} SED fitting.}
    \label{tab:bands}
    \begin{tabular}{ccccc}
        \hline
        Band & Instrument & Wavelength ($\mu$m) & Depth 5$\sigma$ (mag) & Reference \\
        \hline
        $u*$ & CFHT/MegaCam & 0.383 & 27.1 & \cite{Gwyn2012} \\
	$g'$ & CFHT/MegaCam & 0.489 & 27.3 & \cite{Gwyn2012} \\
	$r'$ & CFHT/MegaCam & 0.625 & 27.2 & \cite{Gwyn2012} \\
  	$i'$ & CFHT/MegaCam & 0.769 & 27.0 & \cite{Gwyn2012} \\
        $z'$ & CFHT/MegaCam & 0.888 & 26.1 & \cite{Gwyn2012} \\
        \hline
        F606W & HST/ACS & 0.596 & 28.8 & \cite{Koekemoer2011} \\
        F814W & HST/ACS & 0.809 & 28.2 & \cite{Koekemoer2011} \\
        F125W & HST/WFC3 & 1.250 & 27.6 & \cite{Koekemoer2011} \\
        F140W & HST/WFC3 & 1.397 & 26.8 & \cite{Brammer2012,Skelton2014} \\
        F160W & HST/WFC3 & 1.542 & 27.6 & \cite{Koekemoer2011} \\
        \hline
	$J$ & WIRCam & 1.254 & 24.4 & \cite{Bielby2012} \\
	$H$ & WIRCam & 1.636 & 24.5 & \cite{Bielby2012} \\
	$Ks$ & WIRCam & 2.159 & 24.3 & \cite{Bielby2012} \\
        \hline
        IRAC1 & $Spitzer$ & 3.6 & 21.8 & \cite{Ashby2015} \\
	IRAC2 & $Spitzer$ & 4.5 & 22.4 & \cite{Ashby2015} \\
        \hline
        F770W & \textit{JWST}/MIRI & 7.7 & 25.4  & \cite{Wu2023} \\ 
        F1000W & \textit{JWST}/MIRI & 10.0 & 24.6 & \cite{Wu2023} \\ 
        F1280W & \textit{JWST}/MIRI & 12.8 &  24.0 & \cite{Wu2023} \\ 
        F1500W & \textit{JWST}/MIRI & 15.0 & 23.5 & \cite{Wu2023} \\ 
        F1800W & \textit{JWST}/MIRI & 18.0 & 22.7 & \cite{Wu2023} \\ 
        F2100W & \textit{JWST}/MIRI & 21.0 & 22.1 & \cite{Wu2023} \\
        \hline
    \end{tabular}
\end{table*}
 
\subsection{\texttt{CIGALE} SED fitting}\label{sec:SED model}
We utilise the Code Investigating GALaxy Emission \citep[\texttt{CIGALE},][]{Boquien2019} to fit the spectral energy distribution (SED) of galaxies. In the energy balance principle, the stellar light emitted in UV/optical is absorbed by dust and re-emitted in the mid- to far-IR. \texttt{CIGALE} fits the multi-wavelength band photometric data based on the energy balance principle using various modules \citep[e.g.,][]{Boquien2019, Wang2020, Daryl2021}. 
We utilise \texttt{CIGALE} version 2020.0 to fit the SEDs with models in Table \ref{tab:cigale} such as star formation history (SFH), attenuation law, dust emission, and single stellar population (SSP). 
Important quantities obtained by \texttt{CIGALE} such as the photometric redshift, total infrared luminosity, and star formation rate (SFR) are presented in Table \ref{tab:results}. We utilise the delayed SFH optional exponential burst from \cite{Boquien2019}. We adopt the stellar templates from \cite{Bruzual2003} with the initial mass function \cite{Salpeter1955}. The dust attenuation model follows \cite{Charlot2000} with additional flexibility. The attenuation law model is contributed by the birth cloud and the interstellar medium (ISM) with two power laws \citep{LoFaro2017} and \citep{Buat2018}, and we set slope flexibility. In addition, we parameterised the ISM V-band attenuation ($A^{ISM}_{V}$) and the $\mu = 0.44$ based on \citep{Malek2018}. The dust absorbed from UV/optical and re-emitted into IR is modelled from dust templates of \citep{Draine2014}. Finally, we incorporated the AGN emission model module proposed by \citet{Fritz2006}, which employs a radiative transfer model incorporating three key components: (1) the primary source of radiation situated within the torus, (2) the scattered emission by dust, and (3) the thermal dust emission, as described by \citet{Boquien2019}.  Compared to \citep{Daryl2021}, we tried a slightly larger number of parameters (e.g., 0.1 for optical depth; 0.0 for beta in gas density;  0.0 for gamma in gas density;  60 for opening angle; 0.001 for the angle between equatorial axis and line of sight).
The redshifting module contains the K-correction to estimate the photometric redshift \citep{Boquien2019}. The photometric redshift is a free parameter in the SED fitting with an interval of 0.01 for $z=0-2$, and 0.1 for $z=2-20$.

\begin{table*}
        \centering
        \caption{List of modules and parameter settings for our \texttt{CIGALE} SED fitting.}
        \label{tab:cigale}
        \begin{tabular}{cc}
             \hline
             \textbf{Parameters} & \textbf{Value} \\
             \hline
             \multicolumn{2}{c}{\textit{Delayed SFH with optional exponential burst} \citep{Boquien2019}} \\
             \hline
             e-folding time of the main stellar population [$10^6$ yr] & 2000.0, 5000.0, 10000.0 \\
             Age of the galaxy's main stellar population [$10^6$ yr] & 1000.0, 2000.0, 5000.0, 10000.0 \\
             e-folding time of the late starburst population [$10^6$ yr] & 200.0, 500.0, 2000.0, 20000.0\\
             Age of the late burst [$10^6$ yr] & 10.0, 20.0\\
             Mass fraction of the late burst population & 0.00, 0.01, 0.05, 0.1 \\
             multiplicative factor controlling the amplitude of SFR if normalisation is True & 1.0 \\
             \hline
             \multicolumn{2}{c}{\textit{SSP} \citep{Bruzual2003}} \\
             \hline
             Initial mass function & \cite{Salpeter1955} \\
             Metallicity & 0.02 \\
             Age of separation between the young and old star populations & 10.0 \\
             \hline
             \multicolumn{2}{c}{\textit{Dust attenuation} \citep{Charlot2000}} \\
             \hline
             Logarithm of the V-band attenuation in the ISM &  0.1, 0.17, 0.28, 0.46, 0.77, 1.29, 2.15, 3.59, 5.99, 10.0\\
             Ratio of V-band attenuation from old and young stars & 0.44 \\
             Power-law slope of the attenuation in the ISM & -10.0, -5.0, -1, 
             -0.9,-0.7,-0.5,-0.1 \\
             Power-law slope of the attenuation in the birth cloud & -10.0, -5.0, -3.0, -1.3, -1.0, -0.7\\
             \hline
             \multicolumn{2}{c}{\textit{Dust emission} \citep{Draine2014}} \\
             \hline
             Mass fraction of PAH & 0.47, 1.12, 1.77, 2.5, 3.19, 3.9, 4.58, 5.26, 5.95, 6.63, 7.32 \\
             Minimum radiation field (${\rm U}_{\rm min}$) & 0.1, 1.0, 10.0 \\
             Power-law slope $\alpha$ ($\frac{{\rm dU}}{{\rm dM}} \propto U^\alpha$) & 1.0, 3.0 \\
             Fraction illuminated from ${\rm U}_{\rm min}$ to ${\rm U}_{\rm max}$ & 0.01, 0.05, 0.1 \\
             \hline
             \multicolumn{2}{c}{\textit{AGN emission} \citep{Fritz2006}} \\
             \hline
             Ratio of the maximum and minimum torus radii & 60.0 \\
             Optical depth at 9.7 $\mu$m & 0.1, 0.3,6.0\\
             Value of $\beta$ in gas density gradient along the \\ radial and polar distance coordinates \citep[Eq. 3 in][]{Fritz2006} & 0.0, -0.5 \\
             Value of $\gamma$ in gas density gradient along the \\ radial and polar distance coordinates \citep[Eq. 3 in][]{Fritz2006} & 0.0, 4\\
             Opening angle of the torus & 60.0, 100\\
             Angle between equatorial axis and line of sight & 0.001, 60.1,89.99\\
             AGN contribution fraction (${\rm frac}_{\rm AGN}$) & 0.025 for 0.0 - 0.7 \\
             \hline
             \multicolumn{2}{c}{\textit{Redshifting} \citep{Boquien2019}}\\
             \hline
             Redshift & 0.01 for $z=0-2$, and 0.1 for $z=2-20$ \\
             \hline
        \end{tabular}
\end{table*}

\section{Results}\label{sec:results}
\begin{figure}
    \includegraphics[width=\columnwidth]{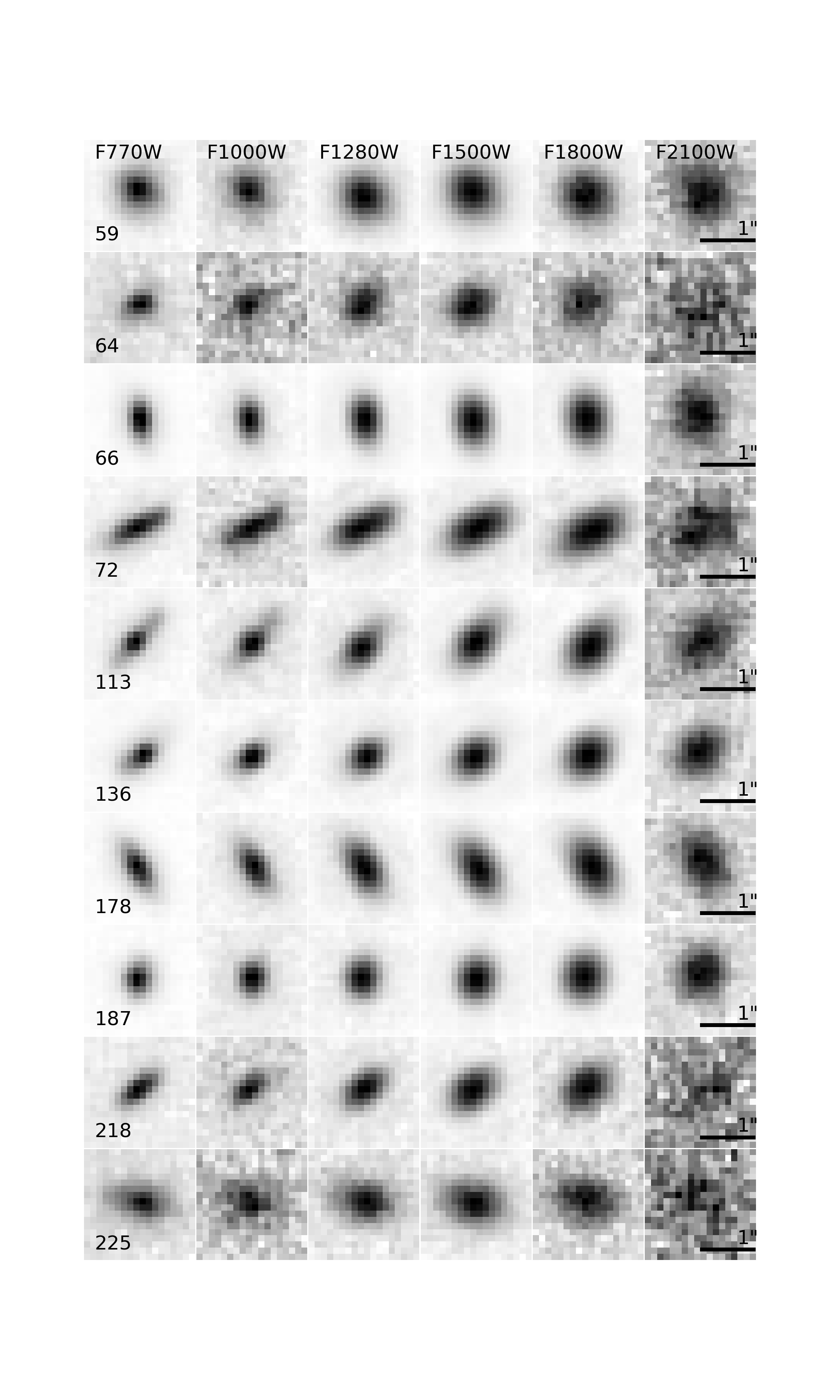}
    \caption{The \textit{JWST} MIRI cutout images ($2"\times2"$) of our sample galaxies selected by the prominent PAH feature in the mid-IR. ID numbers (from Table \ref{tab:results}) are given on the leftmost panel. The pixel scale is 0.11", and a small horizontal bar on the rightmost panel indicates the $1"$ size on the sky.}
    \label{fig:cutout}
\end{figure}

\begin{figure}
	\includegraphics[width=\columnwidth]{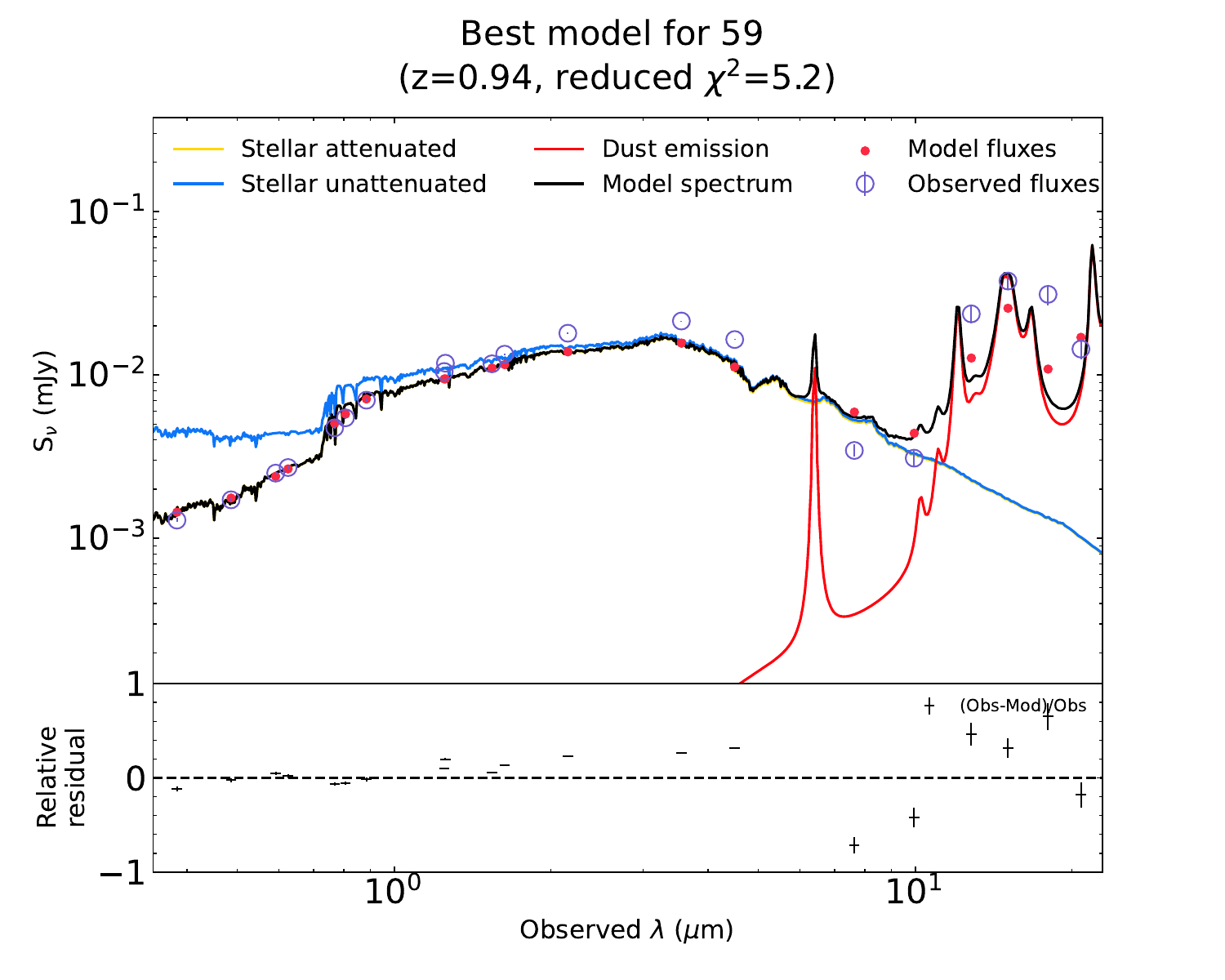}
    \caption{The SED fitting for ID 59 in the Table \ref{tab:results}. The blue curve is the stellar emission and the red curve is the dust emission. The purple points are the observed photometry. The estimation of the photometric redshift was carried out simultaneously.}
    \label{fig:SED example1}
\end{figure}

\begin{figure}
    \includegraphics[width=\columnwidth]{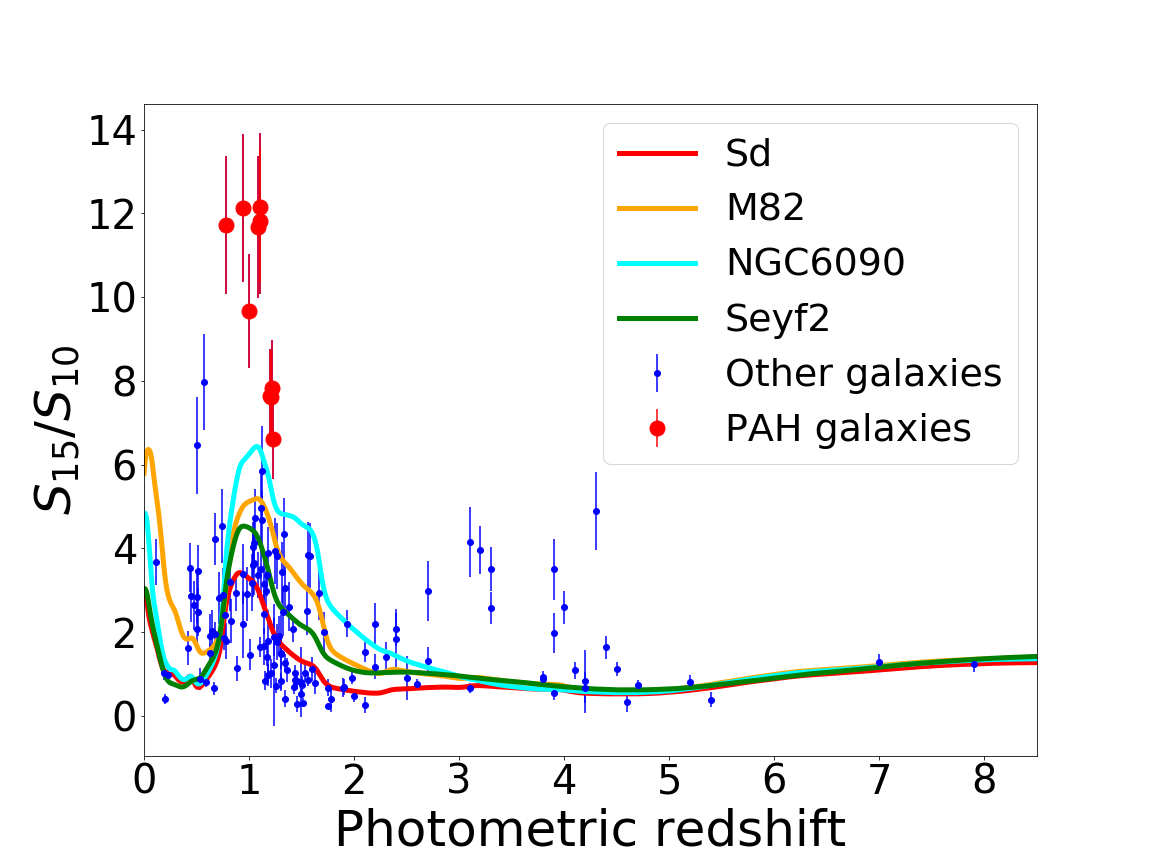}
    \caption{The distribution of the flux density ratio $S_{15}/S_{10}$ for the CEERS MIR galaxies as a function of photometric redshift. Our sample galaxies (selected by "15 $\mu$m excess", $log(S_{15}/S_{10})>0.8$) are shown by the red dots. 
    Curves represent the estimated evolution of $S_{15}/S_{10}$ colour based on the SED model templates \citep{Polletta2007}. Spiral galaxy (Sd), a starburst having moderate PAHs (M82), a starbutst showing stronger PAHs (NGC6090), and Seyfert 2 galaxy are shown. See the legend in the upper right box. 
    }
    \label{fig:redshift vs flux density ratio}
\end{figure}

We identify a subsample of 10 PAH luminous galaxies through SED fitting using \texttt{CIGALE}. These galaxies are selected based on a reduced chi-square ($\chi^2$)  less than 6 to ensure reliable fitting results. Two galaxies which do not meet the criterion are excluded, as shown in Fig. \ref{fig:SED example11}, \ref{fig:SED example12}. The estimated physical properties of the galaxies are listed in Table \ref{tab:results}. The AGN contribution obtained from \texttt{CIGALE} is zero for these ten sources, and ID 66 and 136 have 0.025. Since AGN contribution is very small (less than 2.5\%), we regard dust luminosity from \texttt{CIGALE} as total infrared luminosity. In Fig. \ref{fig:cutout}, we show cutout images ({2"$\times$2"}) at six MIRI bands of our ten PAH-selected galaxies, and the pixel scale is 0.11". These are classified as extended sources with CLASS\_STAR $<0.5$ in the EGS catalogue \citep{Stefanon_2017}. We note that our MIRI imaging does not provide the required resolution to conclude if these galaxies are interacting systems. As reported by \citet[Fig.1]{Huang2023} NIRCam imaging can provide the necessary resolution at $z\sim1-2$ to analyse the detailed morphologies of these galaxies. In Fig. \ref{fig:SED example1}, we present an example (ID59 in the Table \ref{tab:results}) of the SED fitting result from \texttt{CIGALE}. The model spectrum (black solid line) is mainly composed of two primary components: stellar emission (blue dashed line) and dust emission (red solid line). We also analyse the photometric redshift and the reduced chi-square ($\chi^2$). In Fig. \ref{fig:redshift vs flux density ratio}, we show the redshift against the flux density ratio between $S_{15}$ and $S_{10}$ for all sources in the \textit{JWST} CEERS field. 
Using galaxy SED models from \citet{Polletta2007}, we show how $S_{15}/S_{10}$ of typical galaxy types evolve as a function of redshift, presenting a comparison with the observed data points.  The SED model templates used here are a spiral galaxy (Sd) as a normal star-forming galaxy (SFG), M82 as a starburst, NGC6090 (similar to M82 but showing much stronger PAHs), and Seyfert type2 as a composite. It shows that our sample galaxies are located much higher than what we can expect from the galaxy types we know.  The red points, representing the PAH luminous galaxies, are located around $z \sim 1$. As the rest-frame PAH 7.7 $\mu$m features shift to 15.4$ \mu$m at $z = 1$, this is expected by selecting 15$ \mu$m excess sources.

\begin{table*}
    \centering
    \caption{The 10 PAH luminous galaxies SED fitting results}
    \label{tab:results}
    \begin{tabular}{rrrrrrrr}
        \hline
        ID & RA & DEC & reduced chi-square($\chi^2$) & photometric redshift & $L_{IR}$ ($L_{\odot}$) & SFR ($M_{\odot} {\rm yr}^{-1}$) & M ($M_{\odot}$) \\
        \hline
59 & 215.16776 & 53.051688 & 5.17 & 0.94 & 5.11 $\pm$ 3.05 $\times 10^{10}$ & 12.67 $\pm$ 5.63 & 2.86 $\pm$ 0.56 $\times 10^{10}$ \\
64 & 215.158128 & 53.046943 & 1.37 & 1.0 & 20.72 $\pm$ 18.4 $\times 10^{10}$ & 57.95 $\pm$ 53.22 & 0.71 $\pm$ 0.28 $\times 10^{10}$ \\
66 & 215.159489 & 53.048452 & 3.94 & 1.1 & 38.92 $\pm$ 2.0 $\times 10^{10}$ & 49.12 $\pm$ 3.27 & 12.31 $\pm$ 0.74 $\times 10^{10}$ \\
72 & 215.160424 & 53.051557 & 2.75 & 1.1 & 2.99 $\pm$ 0.84 $\times 10^{10}$ & 5.31 $\pm$ 1.31 & 2.53 $\pm$ 0.25 $\times 10^{10}$ \\
113 & 215.16312 & 53.064216 & 4.29 & 1.2 & 11.39 $\pm$ 5.32 $\times 10^{10}$ & 16.45 $\pm$ 14.37 & 3.91 $\pm$ 0.5 $\times 10^{10}$ \\
136 & 215.145024 & 53.058089 & 3.26 & 1.22 & 9.54 $\pm$ 3.53 $\times 10^{10}$ & 15.85 $\pm$ 6.07 & 6.31 $\pm$ 0.79 $\times 10^{10}$ \\
178 & 215.094758 & 52.983213 & 5.05 & 1.23 & 43.5 $\pm$ 12.45 $\times 10^{10}$ & 62.92 $\pm$ 21.31 & 6.7 $\pm$ 1.04 $\times 10^{10}$ \\
187 & 215.075643 & 52.973745 & 1.76 & 1.21 & 15.88 $\pm$ 2.71 $\times 10^{10}$ & 21.09 $\pm$ 4.83 & 6.93 $\pm$ 0.8 $\times 10^{10}$ \\
218 & 215.084489 & 52.990326 & 2.33 & 1.09 & 6.3 $\pm$ 3.69 $\times 10^{10}$ & 10.14 $\pm$ 6.54 & 1.62 $\pm$ 0.5 $\times 10^{10}$ \\
225 & 215.068587 & 52.98334 & 4.75 & 0.78 & 27.04 $\pm$ 5.31 $\times 10^{10}$ & 75.92 $\pm$ 15.49 & 0.79 $\pm$ 0.16 $\times 10^{10}$ \\
\hline
    \end{tabular}
\end{table*}

\begin{figure}
    \includegraphics[width=\columnwidth]{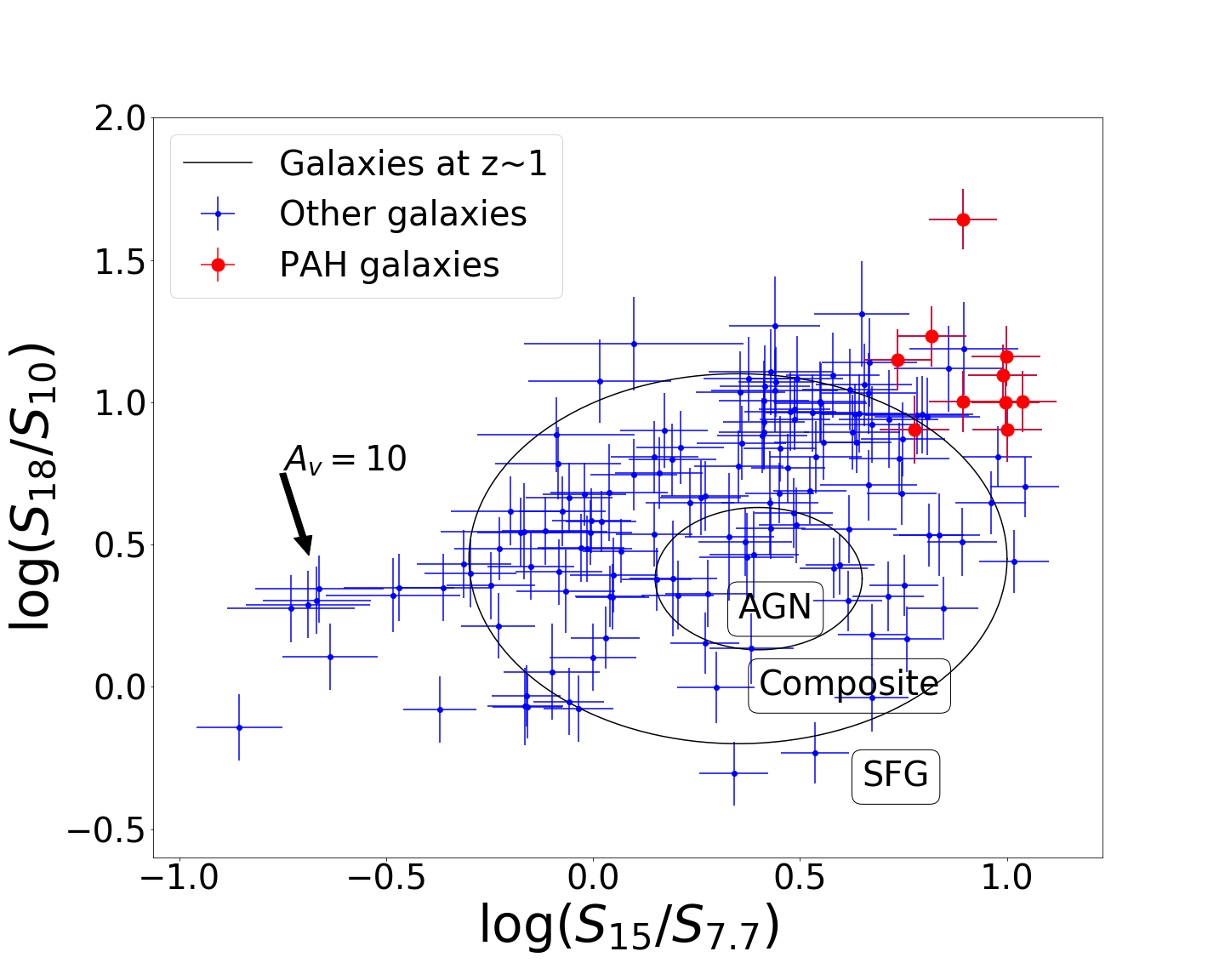}
    \caption{The AGN-Composite-SFG colour selection on $S_{18}/S_{10}$-$S_{15}/S_{7.7}$. The inner circle is the AGN criteria, and the outer circle is the SFG criteria from \citep{Kirkpatrick_2017}. The region between the two circles is composites.  The red points are the PAH luminous galaxies. The black arrow is the extinction $A_{v}=10$.}
    \label{fig:Kirkpatrick2017}
\end{figure}

\section{Discussion}\label{sec:discussion}
\subsection{Colour Selection}\label{sec:colour selection}
\citet{Kirkpatrick_2017} generated the synthetic galaxies from a MIR-based template library \citep{Kirkpatrick2015}. The template library comprises 11 templates derived from spectroscopic sources observed 
 with \textit{Spitzer} and \textit{Herschel}. \citet{Kirkpatrick_2017} utilised 5500 synthetic galaxies to establish \textit{JWST} MIRI colour combination criteria for three distinct populations AGNs, Composite, and SFGs for galaxies within the redshift range of 1 to 2. 
In Fig. \ref{fig:Kirkpatrick2017}, $S_{15}/S_{7.7}$-$S_{18}/S_{10}$ colour-colour plot for \textit{JWST} sources, and two black contours indicate the selection criteria at $z = 0.75 - 1.25$ from  \citet[Eq.1]{Kirkpatrick_2017}. AGNs are located inside the small circle, composites are located between the two circles, and SFGs are located outside the big circle. Red dots represent our ten PAH luminous galaxies, with 9 of 10 sources located in the SFG region, and one source located in the composite region. Note, however, that these boundaries are not clear cuts, but rather a transitional area, where galaxies can mix in and out. None of the PAH galaxies we have selected is in their AGN circle, indicating that our criteria have successfully selected star-forming galaxies without significant contamination from AGN.

\subsection{Obscured AGN or Compact obscured nuclei}
\begin{figure}
	\includegraphics[width=\columnwidth]{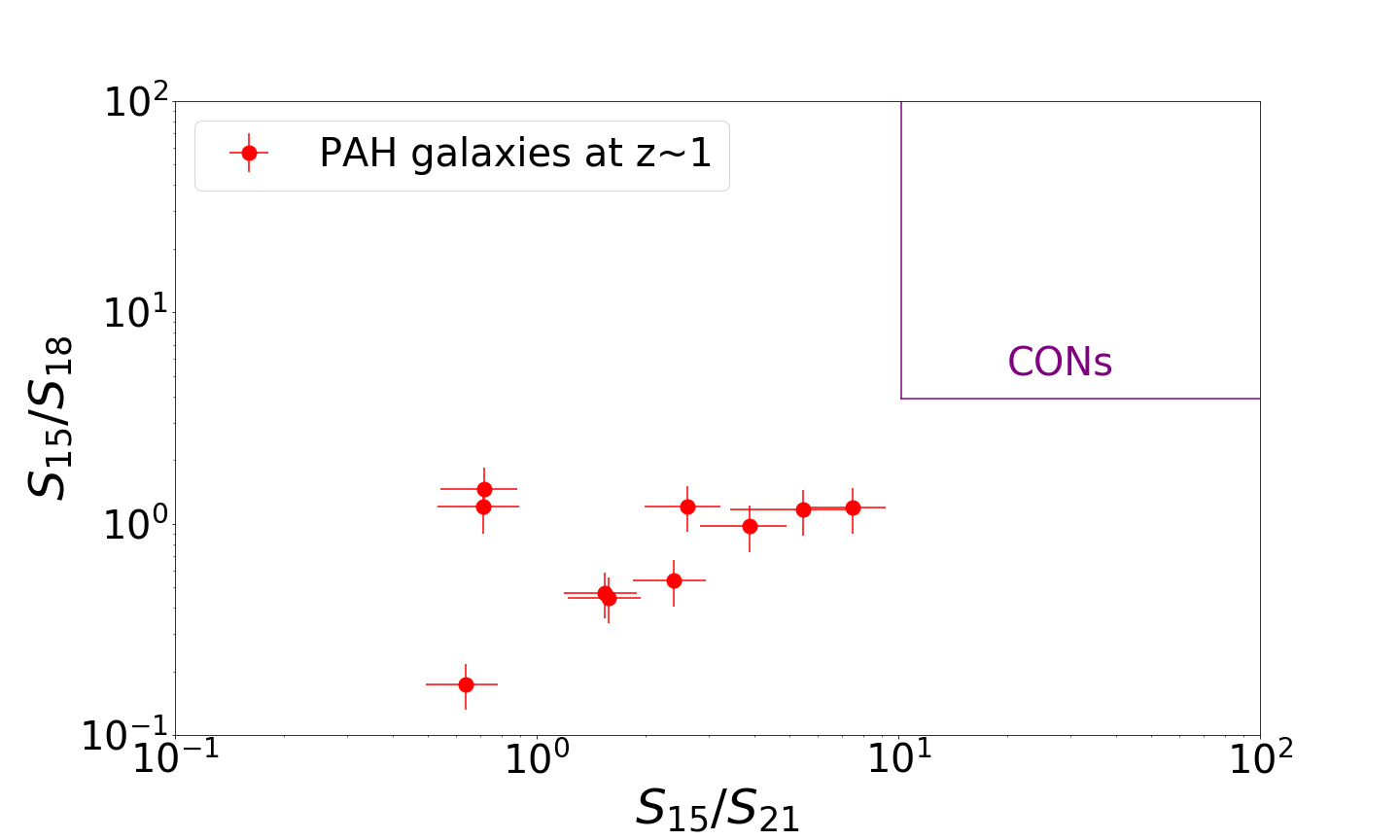}
    \caption{The CON dominated colour selection at z$\sim$1 on $S_{15}/S_{21}-S_{15}/S_{18}$, and the selection is from \citep{Garcia-Bernete2022}. The purple square is the CON-dominated region.}
    \label{fig:CON criteria}
\end{figure}

\cite{Falstad2021} discussed the compact obscured nuclei (CON) with identifications of HCN molecule emission lines in the spectrum. \cite{Garcia-Bernete2022} discussed CONs with the PAH 6.2, 11.3 and 12.7 $\mu$m equivalent width (EW) ratio and 9.7 $\mu$m silicate absorption line, and performed the CONs criteria for the \textit{JWST} MIR photometry observation at redshift from 0 to 1.5. In Fig. \ref{fig:CON criteria}, we show the criteria from \citep{Garcia-Bernete2022} given the CON-dominated regions on the colour-colour plot. We found none of our PAH galaxies are located in the CON region. In Sec.\ref{sec:colour selection}, we compared the distribution of our sample with the colour criteria from \citet{Kirkpatrick_2017} which classify their sample into AGN, composite, and SFG. Where should deep obscured AGN locate on the colour-colour plot? In Fig. \ref{fig:Kirkpatrick2017}, we calculate the dust extinction from \citep{Gordon2023}, and we show the extinction arrow $A_{v}=10$ at $z\sim1$. Due to the 9.7 $\mu$m silicate absorption, $S_{18}/S_{10}$ colour at $z\sim1$ becomes bluer by adding more extinction. If our sources experience heavy dust attenuation, after the correction,  the sources will shift in the top left direction, the opposite of the arrow direction. Therefore, in this colour, the selected PAH galaxies are not consistent with the AGN and the composite even after possible extinction correction, except for one galaxy that is already in the composite region.

\subsection{Emission lines contribution to the F1000W and F1500W filters}
We utilised the 15 $\mu$m excess to select the PAH luminous galaxies. However, in \cite{Garcia-Bernete2022a,Garcia-Bernete2022b,Rich2023} discussed that the emission line gives the contribution to the photometry observation. \cite{Chastenet2023a, Chastenet2023b} discuss the CO, HI, and H$\alpha$ emission lines contribution. \cite{Rich2023} said that the AGN-dominated star-forming region shows extremely low 3.3 and 6.2 $\mu$m PAH equivalent width, and there are strong Pfund $\alpha$ emission lines at 7.46$\mu$m. The excess in the photometry has the probability of contributing to the emission line. The \textit{JWST} F1000W, F1500W filters have bandwidths of $\Delta \lambda $ 1.8, 2.92 $\mu$m, respectively. The half-power wavelengths, where the transmission falls to 50\% of its peak value, range from 9.023 to 10.891$\mu$m and 13.527 to 16.664$\mu$m \citep{Rieke2015}. It is a concern that hydrogen emission lines might contribute to the flux in these filters, while our main focus is to estimate the effect of the PAH emissions. For galaxies at $z\sim1$, five hydrogen emission lines fall into F1000W or F1500W filter broadband. In the F1500W filter at $z\sim1$, two emission lines are observed, Pfund $\alpha$ at 7.46 $\mu$m, and Humphreys $\beta$ at 7.49 $\mu$m. In the F1000W filter,  three emission lines could contribute, Pfund $\beta$ at 4.65 $\mu$m, and Humphreys (10-6) and (11-6) at 5.13 and 4.67 $\mu$m. We estimate the contributions by utilising the SFR calculated by \texttt{CIGALE} as listed in Table \ref{tab:results}. Following the \cite{Kennicutt1998} Eq.2, we determined the $L_{H\alpha}$. We estimate contributions from the Pfund $\alpha$ and Humphreys $\beta$, assuming the CASE B recombination at $T=10000K$, $N_{e}= 10000 cm^{-3}$ \citep{Hummer&Storey1987}. Our calculations revealed that the emission lines in the F1000W filter have contributions ranging from 0.01-2.4\% for Pfund $\beta$ and 0.006-1.15\%, 0.005-0.88\%  for Humphreys (10-6), (11-6), respectively. In the F1500W filter, the contributions are 0.006-0.7\% for Pfund $\alpha$, and 0.001-0.2\% for Humphreys $\beta$. Overall, the hydrogen recombination emission lines contribute less than 5\% to the \textit{JWST} F1000W, F1500W filters.

\subsection{Star Formation Main Sequence}
\begin{figure}
	\includegraphics[width=\columnwidth]{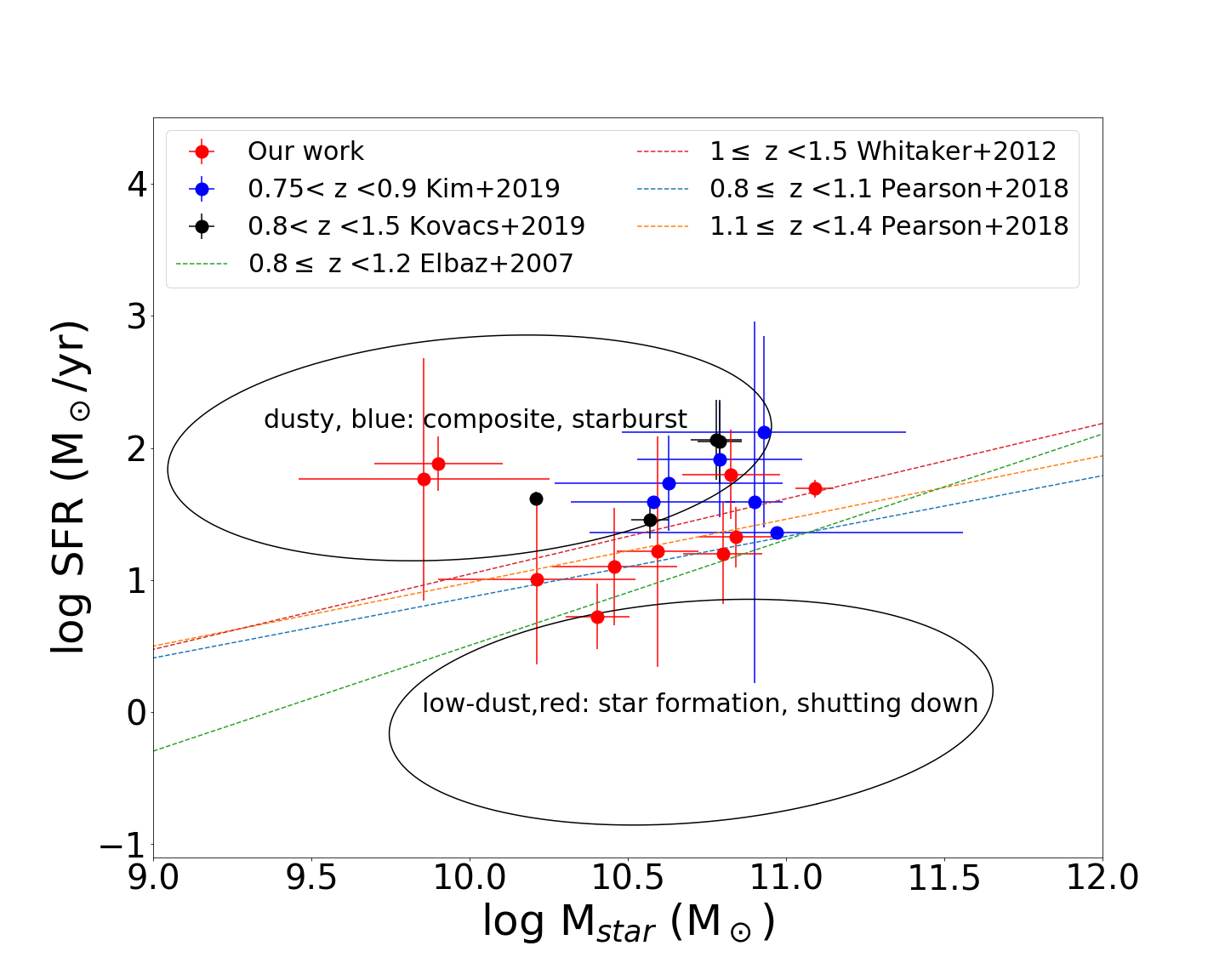}
    \caption{SFR - M$_{\odot}$ correlation at z $\sim$ 1 . Both SFR and stellar mass are calculated from \texttt{CIGALE}. The dashed lines are the results with different redshift bins. The blue, and yellow lines are $z=0.8-1.1$, $z=1.1-1.4$ from \citep{Pearson2018}. The green line is $z=0.8-1.2$ from \citep{Elbaz2007}, and the red line is $z=1-1.5$ from \citep{Whitaker2012}. Blue dots are from \citep{Kim2019}, and the black dots are from \citep{Kovacs2019}. All of the points are located nearby the main sequence.}
    \label{fig:main sequence}
\end{figure}

The main sequence (MS) is a correlation between SFR and stellar mass (M) of star-forming galaxies. MS evolution as a function of redshift is shown and discussed recently \citep{Elbaz2007,Whitaker2012,Pearson2018,Popesso2023}. We calculated the SFR and M from \texttt{CIGALE}, listed in Table \ref{tab:results}, and compared the results at the similar redshift bin with literature. In Fig. \ref{fig:main sequence}, \citet{Pearson2018} gives MS in $z=0.8-1.1$, $z=1.1-1.4$, and \citet{Elbaz2007} gives $z=0.8-1.2$. \citet{Whitaker2012} gives $z=1-1.5$, and the outlier regions with other type galaxies. We over-plot four galaxies between $0.8<z<1.5$ from the \textit{AKARI} catalogue in \citep{Kovacs2019}. 
Our samples plotted in red are mostly located in the main sequence, except for two galaxies in the upper left region as starbursts or composite. One of the two objects is the composite in Fig. \ref{fig:Kirkpatrick2017}. However, both have the AGN fraction=0 in \texttt{CIGALE} SED fitting. Note that \texttt{CIGALE} SED fitting utilises 21 available filters, in contrast to the colour-colour diagram which relies on only 4 filters. Therefore, we regard the \texttt{CIGALE} SED fitting as more informative. It is promising that even though Fig. \ref{fig:Kirkpatrick2017} only utilises 4 colours, we have obtained consistent results from the remaining sources. Compared to previous data from the \textit{AKARI}, and \textit{Spitzer} telescope, \textit{JWST} observed galaxies with less massive and smaller total IR luminosity.

\subsection{PAH luminosity}
\begin{figure}
    \includegraphics[width=\columnwidth]{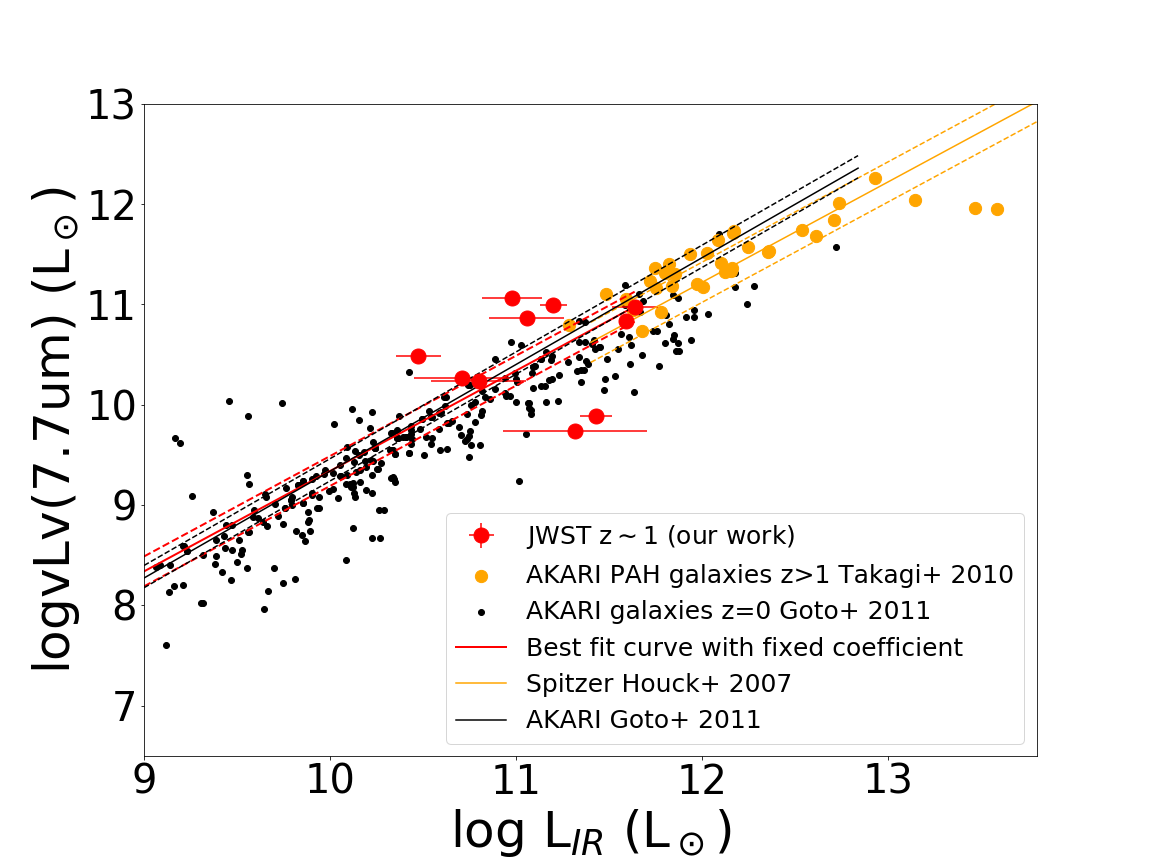}
    \caption{The 7.7 $\mu$m PAH peak luminosity versus total infrared luminosity. The red dots are our work based on \textit{JWST}/MIRI. The orange dots are the \textit{AKARI} results at $z > 1 $ from \citep{Takagi2010}. The black dots are the results from \textit{Spitzer} 8 $\mu$m at the local universe from \citep{Goto2011}. The black solid line was given in \citep{Houck2007}, which indicates the relation for starbursts and two black dashed lines were one sigma dispersion. The red solid line used the same coefficient in our data, the line located within 1 sigma deviation from the orange solid line. The red solid line is the least square fit for our work.}
    \label{fig:L pah vs L IR}
\end{figure}

At a redshift of z $\sim$ 1, the PAH 7.7 $\mu$m feature in galaxies shifts to 15.4 $\mu$m, leading to a 15$\mu$m-excess, which we used to identify PAH luminous galaxies \citep{Takagi2010}. To quantify the strength of PAH features in these galaxies, we employed a relation between PAH peak luminosity and total infrared luminosity for starburst, as established by \citep{Houck2007,Weedman2008}. The equation is given as,
\begin{equation}\label{eq:Houck}
\mathrm{log}(vL_{v}(7.7 \mu m)) = \mathrm{log}(L_{IR}) - 0.78(\pm 0.2),
\end{equation}
$vL_{v}(7.7 \mu m)$ is the PAH 7.7 $\mu$m peak luminosity which indicates the strength of PAH features in the galaxies, and $L_{IR}$ is total IR luminosity. In our work, we computed $L_{IR}$ from \texttt{CIGALE} SED fitting results shown in Table \ref{tab:results}. We follow \cite{Houck2007, Kennicutt1998} to estimate the PAH 7.7 $\mu$m peak luminosity at rest frame from \texttt{CIGALE} SED fitting results, the equation given as,
\begin{equation}\label{eq:L_PAH}
vL_{v}(7.7 \mu m) =\frac{c}{\lambda_0}(1+z) \times \frac{4\pi d_L^2}{1+z} f_{v_0}=S_{v}(15 \mu m) \times 9 \times 10^7 \times D_L^2 \div 7.7
\end{equation}
$\lambda_0$ is the rest-frame wavelength, here is 7.7 $\mu$m. $f_{v_0}$ is the observed flux density, here is 15 $\mu$m flux density. $c$ is light speed. $z$ is a photometric redshift. $S_v$ is the flux density of the PAH 15 $\mu$m peak value in the black line, shown in Fig. \ref{fig:SED example1}, and $D_L$ is the luminosity distance calculated from photometric redshift within the cosmological model. \citet{Takagi2010} provided PAH-selected galaxies, which are LIRGs/ULIRGs, the orange points shown in Fig. \ref{fig:L pah vs L IR}, out of all the galaxies at z $>1$ in the \textit{AKARI} catalogue. In Fig. \ref{fig:L pah vs L IR}, we have the galaxies which have the one order fainter $L_{IR}$ than those in the \textit{AKARI} sample from \citep{Takagi2010}. We performed a least square fit in our work with the same coefficient, and the result is given as,
\begin{equation}\label{eq:curve fit}
\mathrm{log}(vL_{v}(7.7 \mu m)) = \mathrm{log}(L_{IR})-0.66(\pm 0.15).
\end{equation}
Compared with \citep{Houck2007, Weedman2008} and \citep{Takagi2010} observed galaxies by \textit{AKARI} at $z\sim$ 1, our result(red solid line) is located within 1-sigma(orange dashed line). Therefore, statistically, the correlations are consistent between our work and literature on the $L_{PAH}$ to $L_{IR}$ plane. 
In Fig. \ref{fig:L pah vs L IR}, we also compare with the local IR galaxies which are shown in black points and $L_{PAH}$ to $L_{IR}$ relation \citep{Goto2011}. Compared with all galaxies in our work, there are no significant differences between the two relations, which indicates the $L_{PAH}$ to $L_{IR}$ relation does not change significantly in these redshift and luminosity ranges.

\begin{figure}
    \includegraphics[width=\columnwidth]{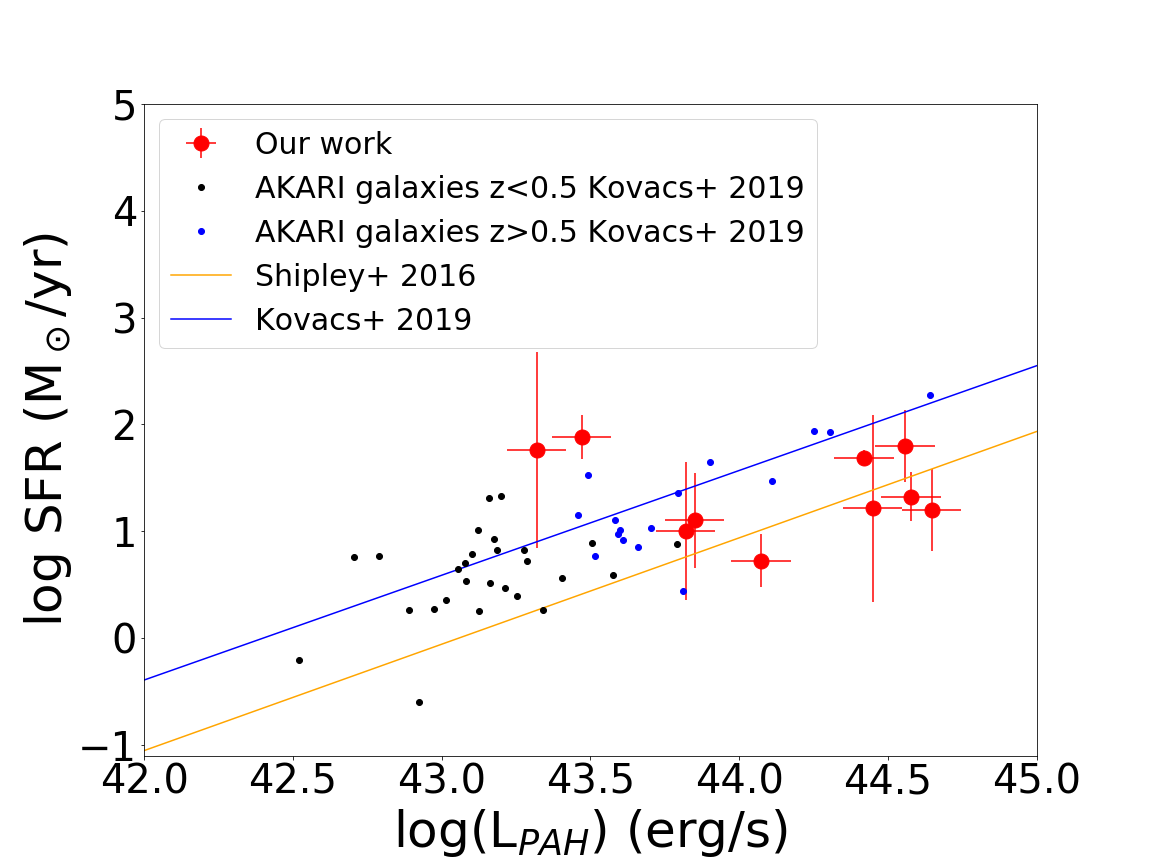}
    \caption{Correlation between PAH peak luminosity and SFR.  Blue and black dots are \textit{AKARI} sources from \citep{Kovacs2019}. Blue solid line is the result from the \textit{AKARI} telescope in Eq. \ref{eq:SFR-PAH2} from \citep{Kovacs2019}, and yellow solid line is the result from \textit{Spitzer} in Eq. \ref{eq:SFR-PAH1} from \citep{Shipley2016}.}
    \label{fig:L pah vs SFR}
\end{figure}

Since PAH is a good tracer of the star-forming region in the Universe, there is a relation between PAH luminosity and SFR, which in turn relates SFR and $L_{PAH}$ for SFGs. Numerous studies support this relationship, as evidenced by galaxy samples and observations at high redshift \citep{Pope2008, Riechers2014, Shipley2016, Spilker2023}.  \cite{Pope2008} presented PAH luminosity as a proxy for the SFR in the submillimeter galaxies (SMGs) from redshift 1 to 2.5 and \cite{Riechers2014} presented one SMG at $z=4.055$ have a consistent correlation between strong PAH features and the SFR. In recent \textit{JWST} observation results, the galaxy at $z=4.2248$ also presents strong PAH emission lines \citep{Spilker2023}. \citet{Shipley2016} gave a relation between the PAH peak luminosity and the star formation rate from \textit{Spitzer} observations as follows:
\begin{equation}\label{eq:SFR-PAH1}
\mathrm{log}(SFR) = -42.84 + 0.995 \times \mathrm{log}(L_{PAH}),
\end{equation}
and \citet{Kovacs2019} gives a correlation from \textit{AKARI}:
\begin{equation}\label{eq:SFR-PAH2}
\mathrm{log}(SFR) = -41.55 + 0.98 \times \mathrm{log}(L_{PAH}).
\end{equation}
$L_{PAH}$ is the PAH peak luminosity $vL_{v}(7.7 \mu m)$ in Eq. \ref{eq:L_PAH} from \citep{Kovacs2019}, the red points are our work in which the SFRs are calculated by \texttt{CIGALE} SED fitting. In Fig. \ref{fig:L pah vs SFR}, we present the $L_{PAH}$ and SFR calculated from \texttt{CIGALE} SED fitting for PAH luminous galaxies.  In our work, the relation suggests a correlation between PAH luminosity and SFR, and the results are similar to Eq. \ref{eq:SFR-PAH2} from \citep{Shipley2016}. However, there is no clear difference for the lower $M$ and fainter $L_{IR}$ galaxies at redshift $\sim$1 compared to previous literature.


\section{Conclusions}\label{sec:conclusions}
We utilise the \textit{JWST} MIRI multi-wavelength photometry and the CANDELS EGS survey to select less luminous infrared galaxies that display prominent MIR PAHs beyond the reach of previous IR telescopes. We discover 10 PAH luminous galaxies out of a sample of 155 galaxies in the two CEERS fields with $log(S_{15}/S_{10})>0.8$. We perform SED fitting on these galaxies using \texttt{CIGALE} models and 21 broad-band photometry from optical to mid-infrared. All of them have a good SED fitting result with a reduced chi-square ($\chi^2$) < 6. These galaxies are located at z $\sim$ 1, and four of them are $L_{IR}$ < 10$^{11}$ $L_{\odot}$. All the properties of the selected PAH galaxies suggest that they are normal star-forming galaxies, rather than U/LIRGs. Our findings are summarised below.
\begin{itemize}
    \item PAH galaxies we selected at redshift $\sim$ 1 in Table \ref{tab:results} have infrared luminosity lower than $L_{IR}$ < 10$^{12}$. \textit{JWST} provides us with the opportunity to investigate faint/normal PAH galaxies, rather than ULIRGs/LIRGs at z$\sim$1.
    \item The selected PAH galaxies fall into the star-forming regions according to criteria from \cite{Kirkpatrick_2017} in the $S_{15}/S_{7.7}$-$S_{18}/S_{10}$ colour space.
    \item The selected PAH galaxies lie on the SF main sequence at z $\sim$ 1, with a stark contrast to ULIRGs, which often lie above the main sequence.
    \item The $L_{PAH}$ versus $L_{IR}$ in Fig. \ref{fig:L pah vs L IR} shows a consistent relation with the known work for ULIRGs. This may indicate that the redshift and luminosity do not impact the PAH luminosity and strength. However, it is important to use a statistical sample to make the conclusion.

    As we have shown a glimpse, the JWST has opened a new window to investigate faint/normal PAH galaxies at z $\sim$ 1. With more \textit{JWST} observations in the near future, the nature of PAH emissions in normal star-forming galaxies will be revealed further.

\end{itemize}
\section*{Acknowledgements}

The authors express their gratitude to the anonymous referee for providing numerous constructive comments that have significantly enhanced the quality of this paper. TG acknowledges the support of the National Science and Technology Council of Taiwan through grants 108-2628-M-007-004-MY3, 111-2112-M-007 -021. and 112-2112-M-007
-013, and 112-2123-M-001 -004.
TH acknowledges the support of the National Science and Technology Council of Taiwan through grants 110-2112-M-005-013-MY3, 110-2112-M-007-034-, and 112-2123-M-001 -004 -.
SH acknowledges the support of The Australian Research Council Centre of Excellence for Gravitational Wave Discovery (OzGrav) and the Australian Research Council Centre of Excellence for All Sky Astrophysics in 3 Dimensions (ASTRO 3D), through project number CE17010000 and CE170100013, respectively.
This work used high-performance computing facilities operated by the Center for Informatics and Computation in Astronomy (CICA) at National Tsing Hua University. This equipment was funded by the Ministry of Education of Taiwan, the National Science and Technology Council of Taiwan, and National Tsing Hua University.
\section*{Data Availability}

Early Release Observations obtained by \textit{JWST} MIRI are publicly available at \href{https://www.stsci.edu/jwst/science-execution/approved-programs/webb-first-image-observations}{https://www.stsci.edu/jwst/science-execution/approved-programs/webb-first-image-observations}.



\bibliographystyle{mnras}
\bibliography{JWST_PAH} 




\appendix

\section{SED result}
\begin{figure}
	\includegraphics[width=\columnwidth]{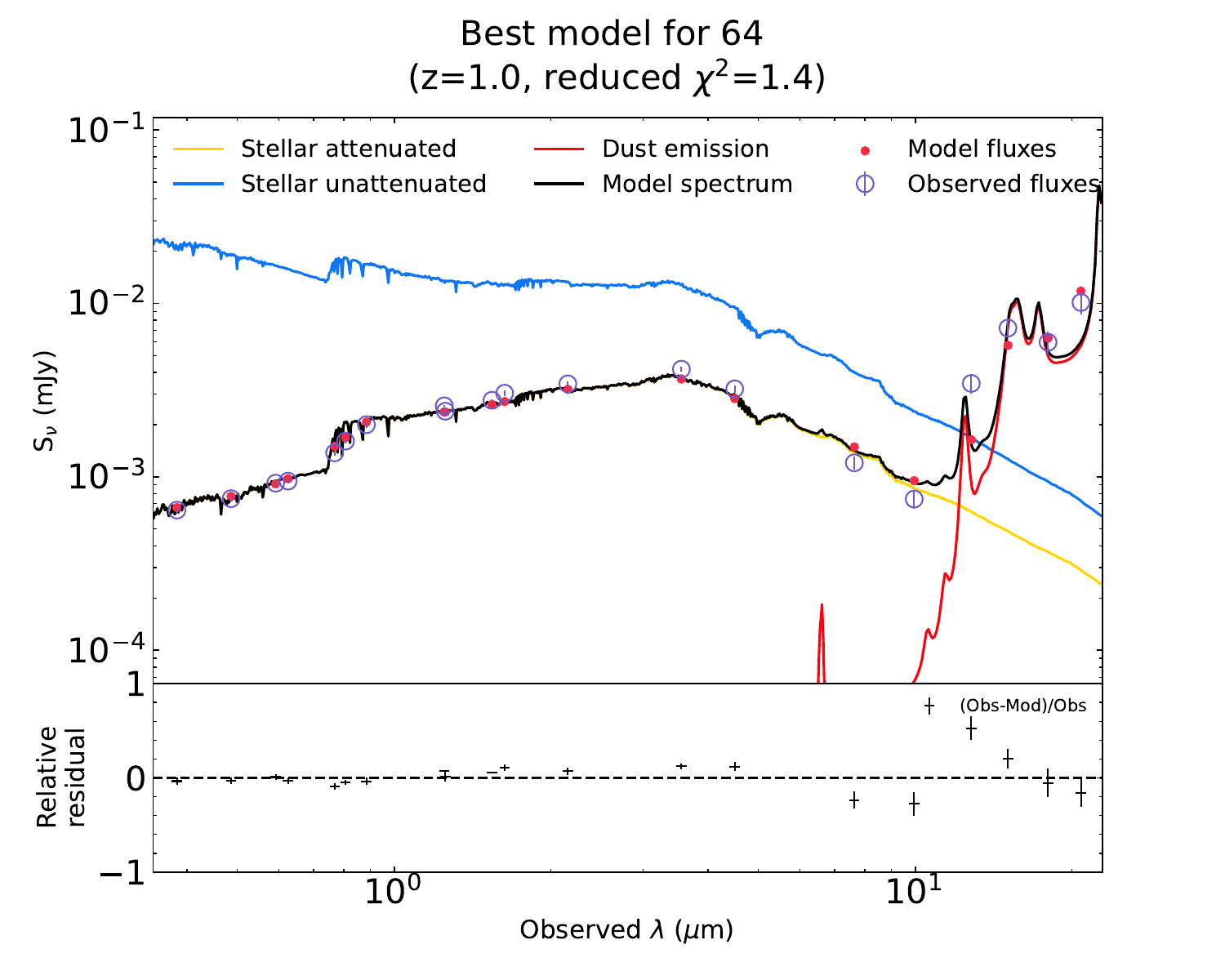}
    \caption{The same as Fig. \ref{fig:SED example1}, but for ID 64}
    \label{fig:SED example2}
\end{figure}

\begin{figure}
	\includegraphics[width=\columnwidth]{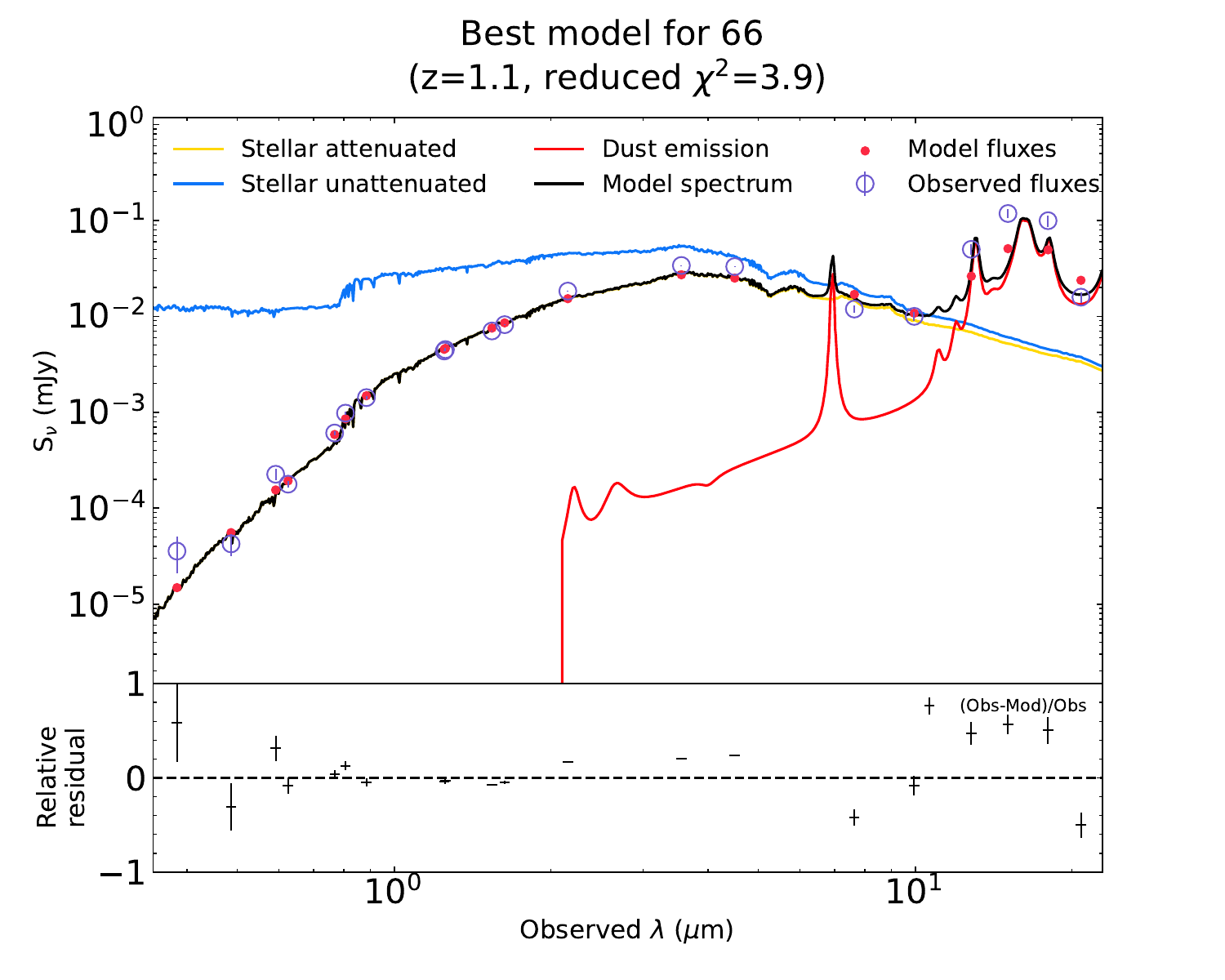}
    \caption{The same as Fig. \ref{fig:SED example1}, but for ID 66}
    \label{fig:SED example3}
\end{figure}

\begin{figure}
	\includegraphics[width=\columnwidth]{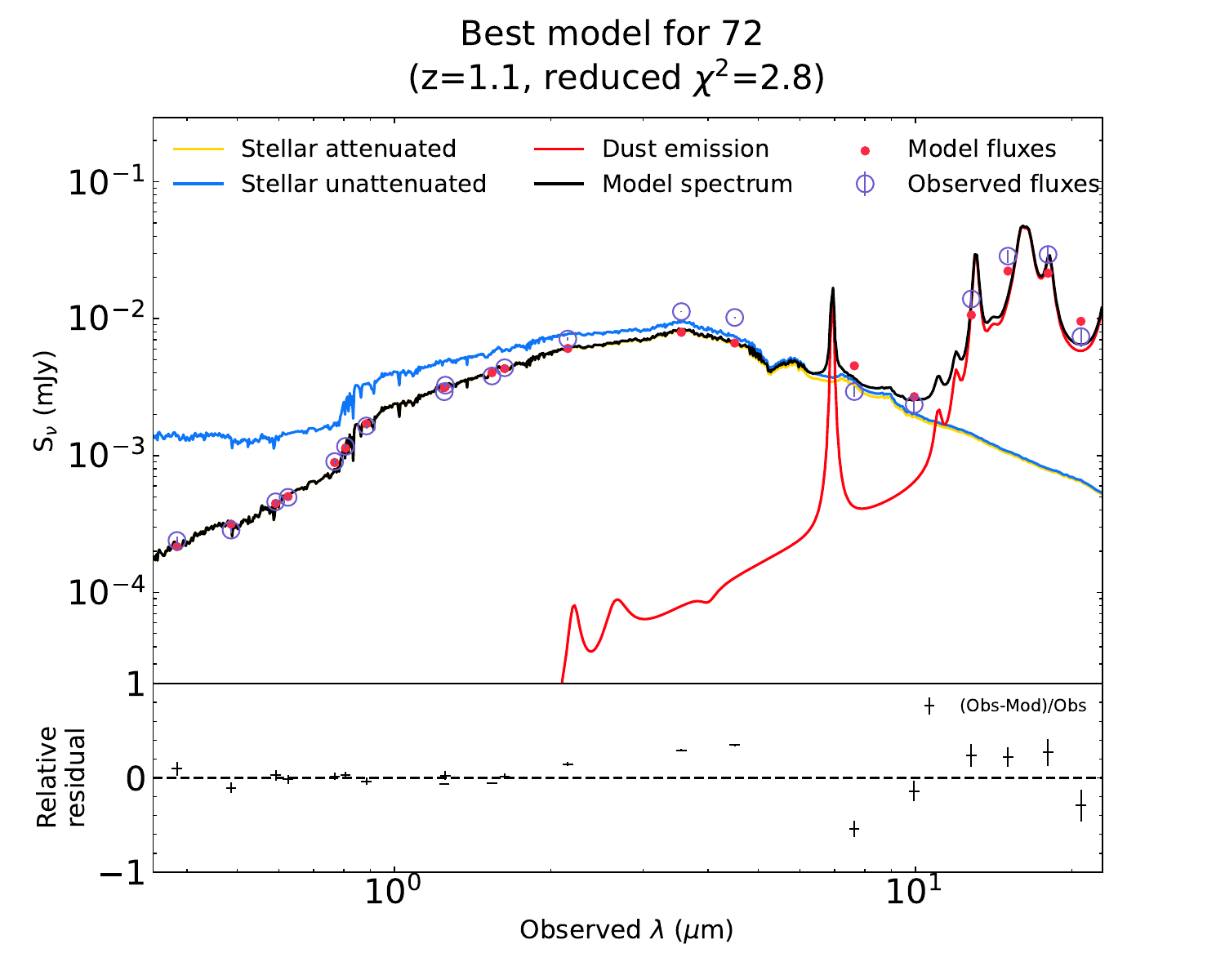}
    \caption{The same as Fig. \ref{fig:SED example1}, but for ID 72}
    \label{fig:SED example4}
\end{figure}

\begin{figure}
	\includegraphics[width=\columnwidth]{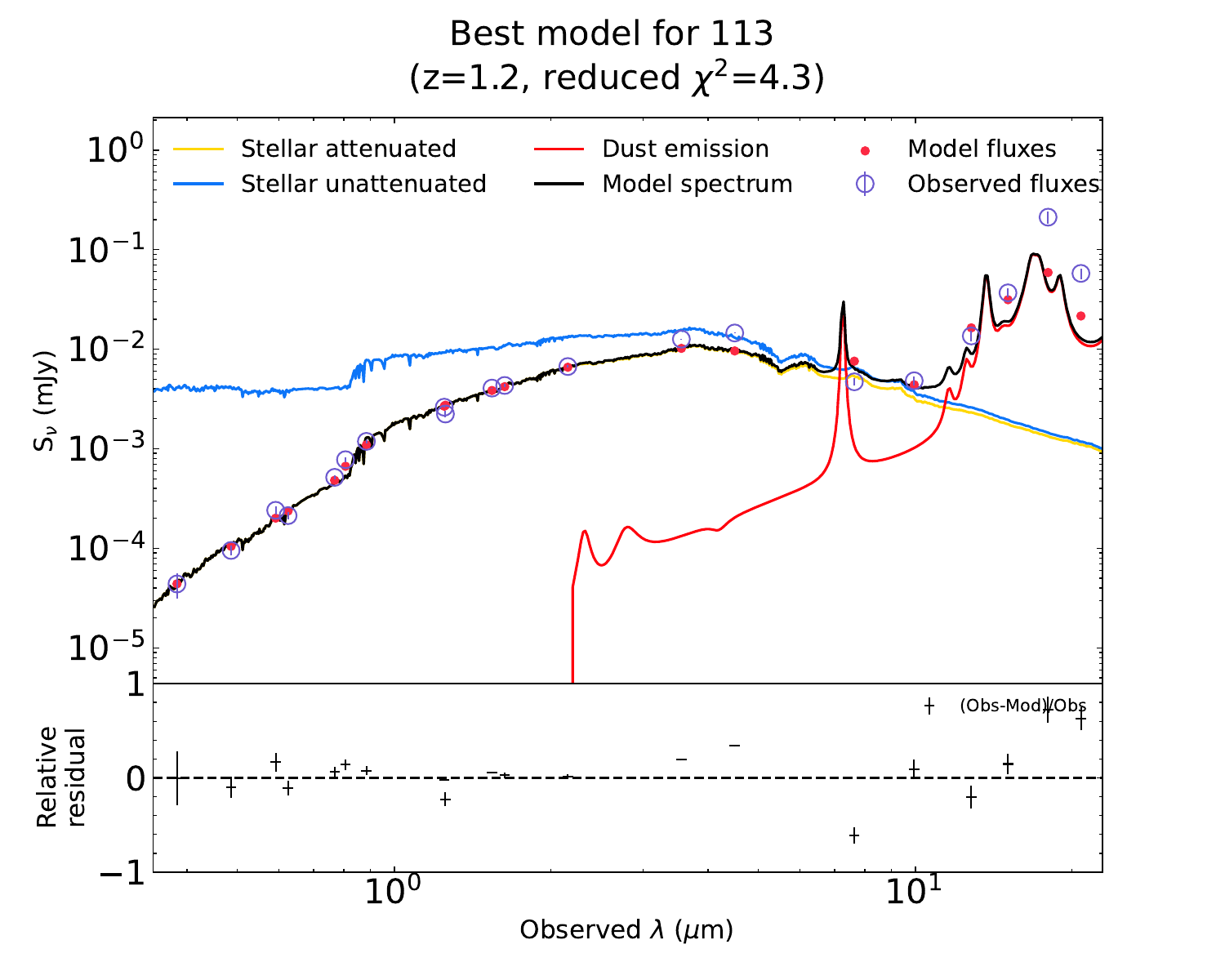}
    \caption{The same as Fig. \ref{fig:SED example1}, but for ID 113.}
    \label{fig:SED example5}
\end{figure}

\begin{figure}
	\includegraphics[width=\columnwidth]{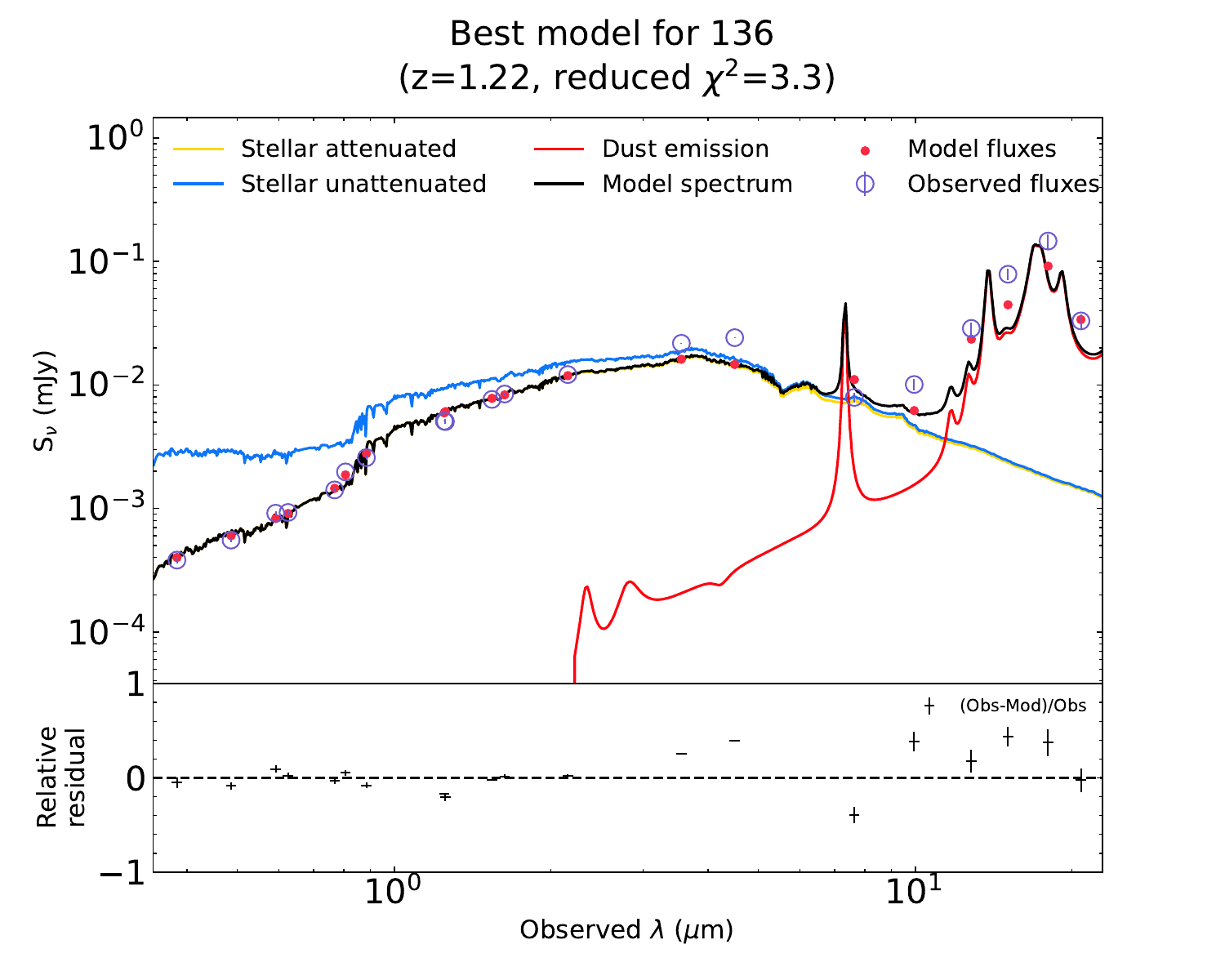}
    \caption{The same as Fig. \ref{fig:SED example1}, but for ID 136.}
    \label{fig:SED example6}
\end{figure}

\begin{figure}
	\includegraphics[width=\columnwidth]{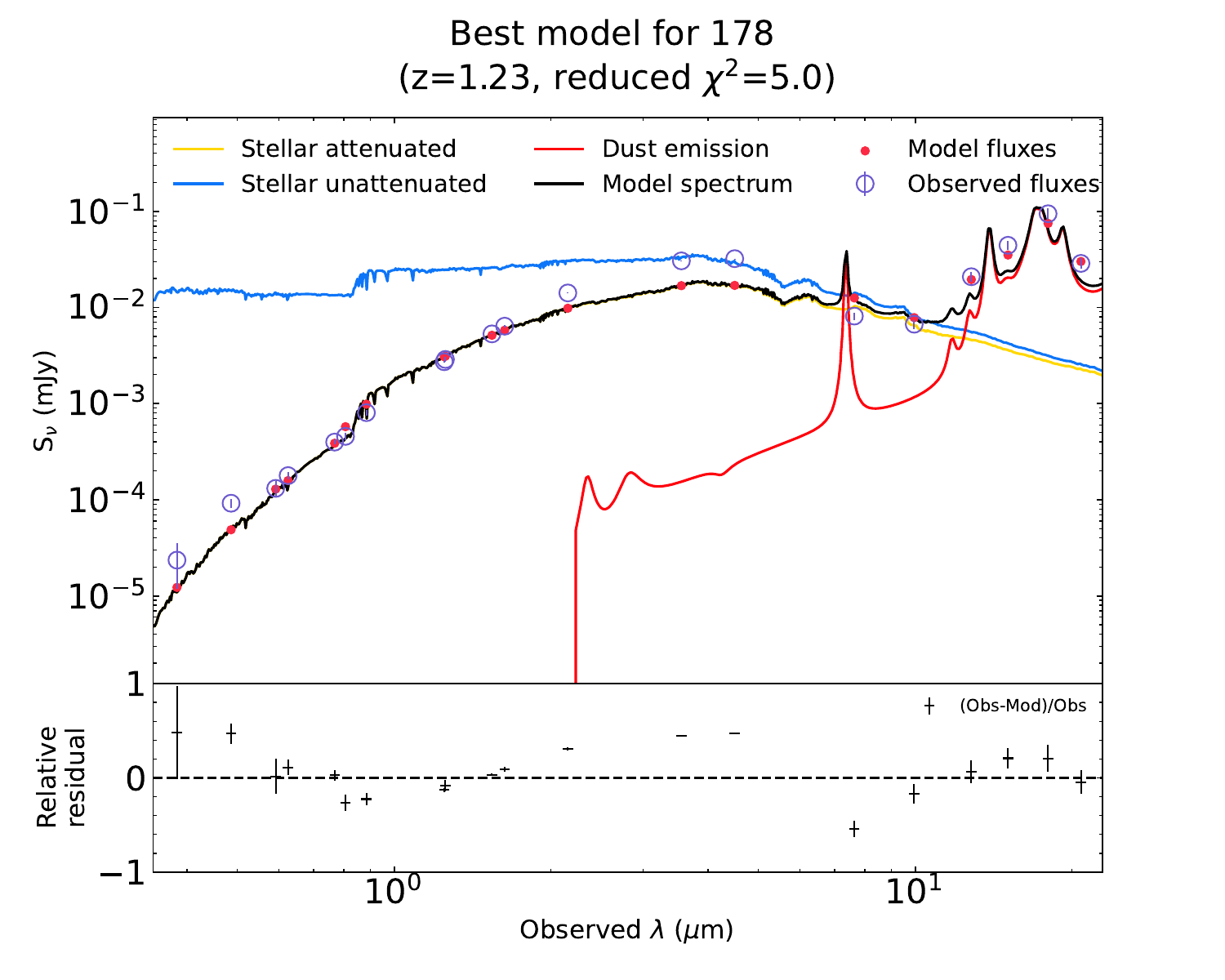}
    \caption{The same as Fig. \ref{fig:SED example1}, but for ID 178.}
    \label{fig:SED example7}
\end{figure}

\begin{figure}
	\includegraphics[width=\columnwidth]{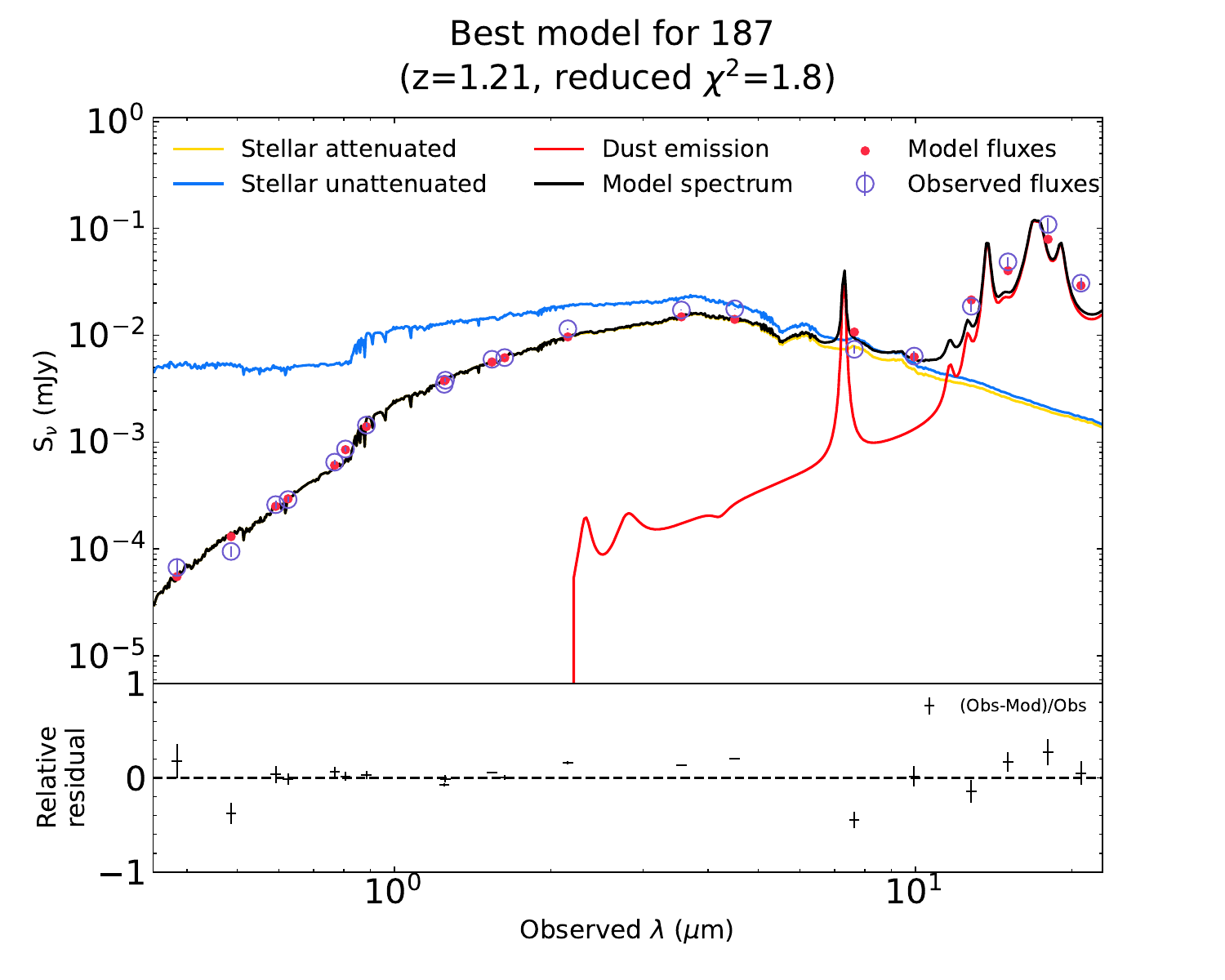}
    \caption{The same as Fig. \ref{fig:SED example1}, but for ID 187.}
    \label{fig:SED example8}
\end{figure}

\begin{figure}
	\includegraphics[width=\columnwidth]{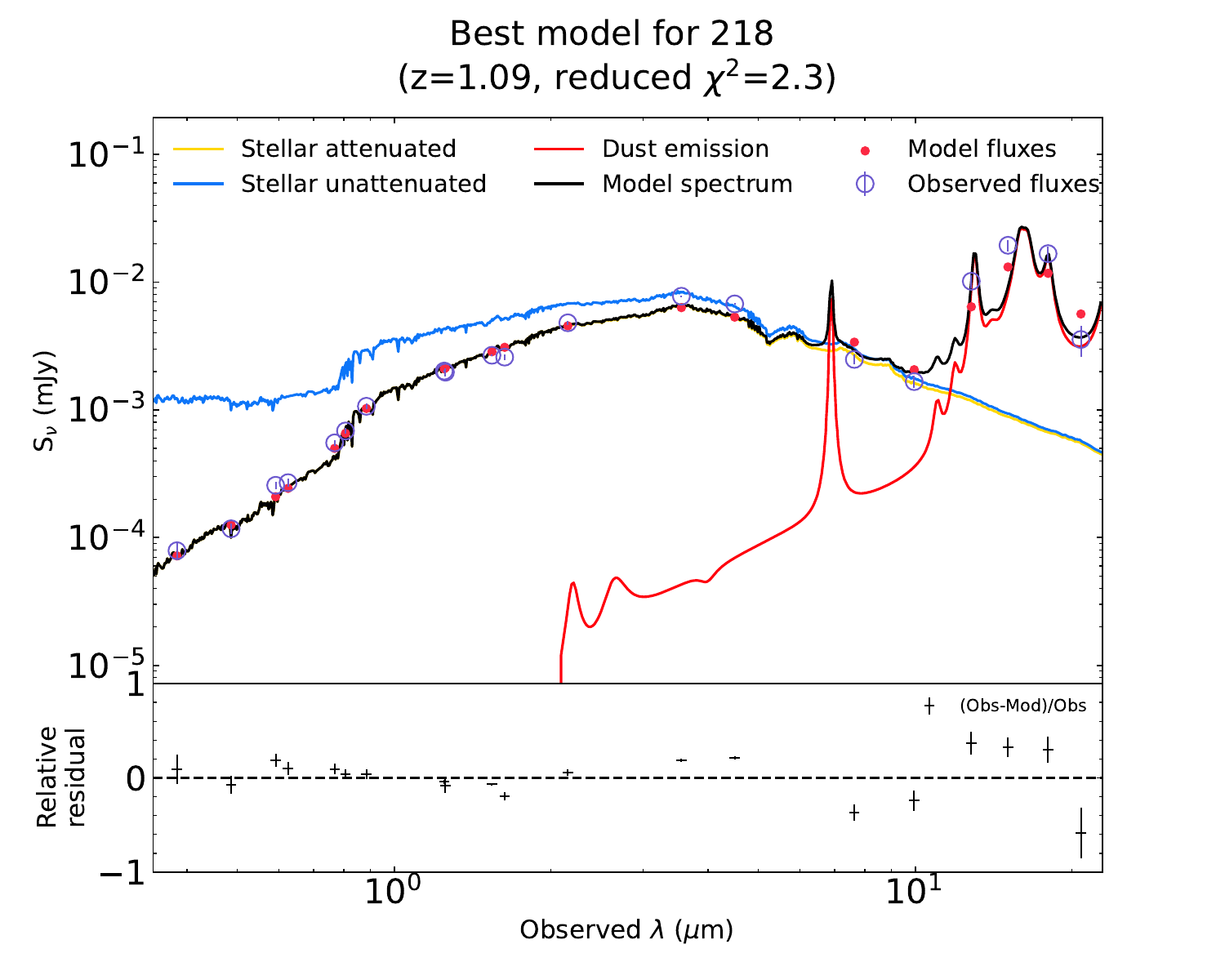}
    \caption{The same as Fig. \ref{fig:SED example1}, but for ID 218.}
    \label{fig:SED example9}
\end{figure}

\begin{figure}
	\includegraphics[width=\columnwidth]{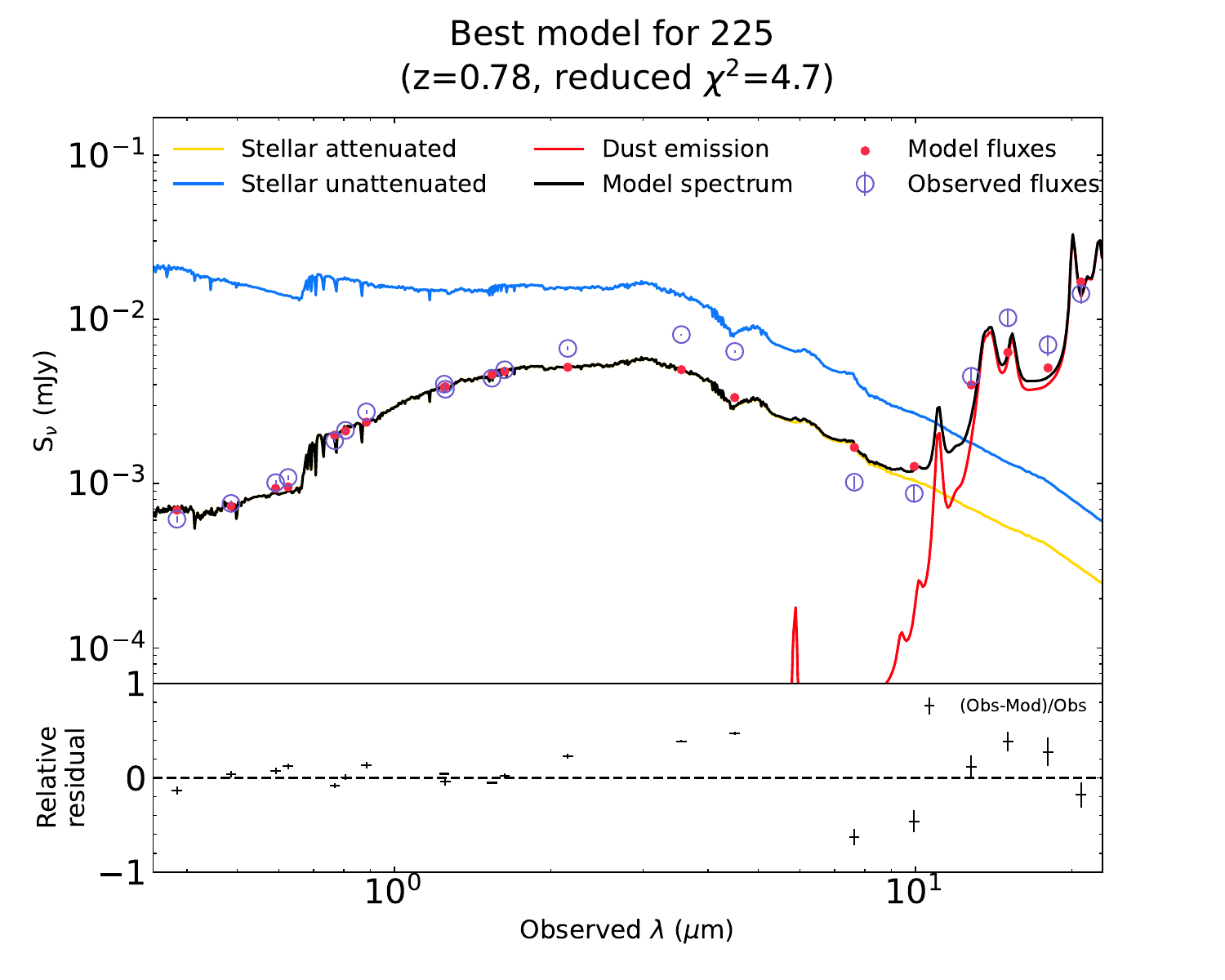}
    \caption{The same as Fig. \ref{fig:SED example1}, but for ID 225.}
    \label{fig:SED example10}
\end{figure}

\begin{figure}
	\includegraphics[width=\columnwidth]{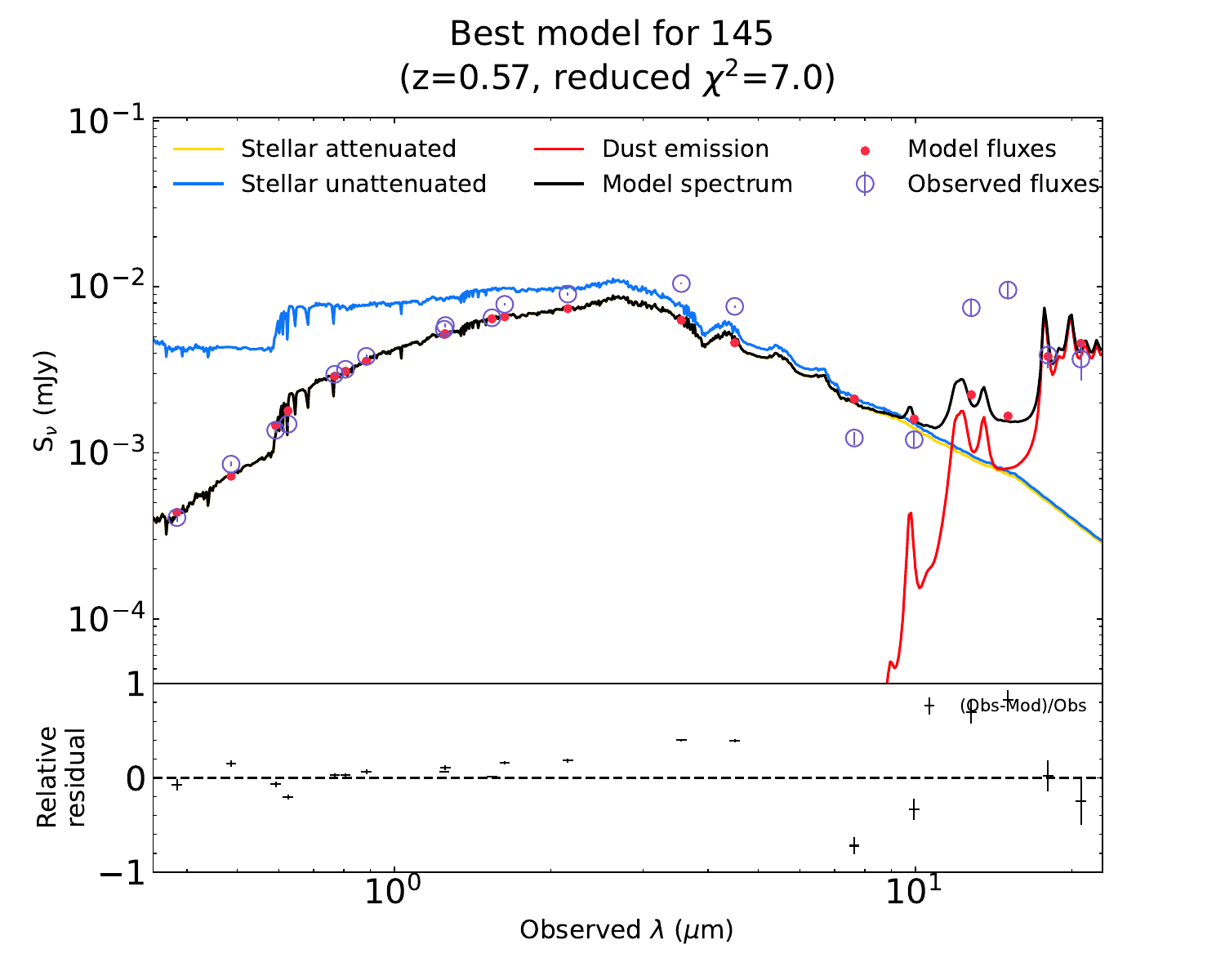}
    \caption{The same as Fig. \ref{fig:SED example1}, but for ID 145.}
    \label{fig:SED example11}
\end{figure}

\begin{figure}
	\includegraphics[width=\columnwidth]{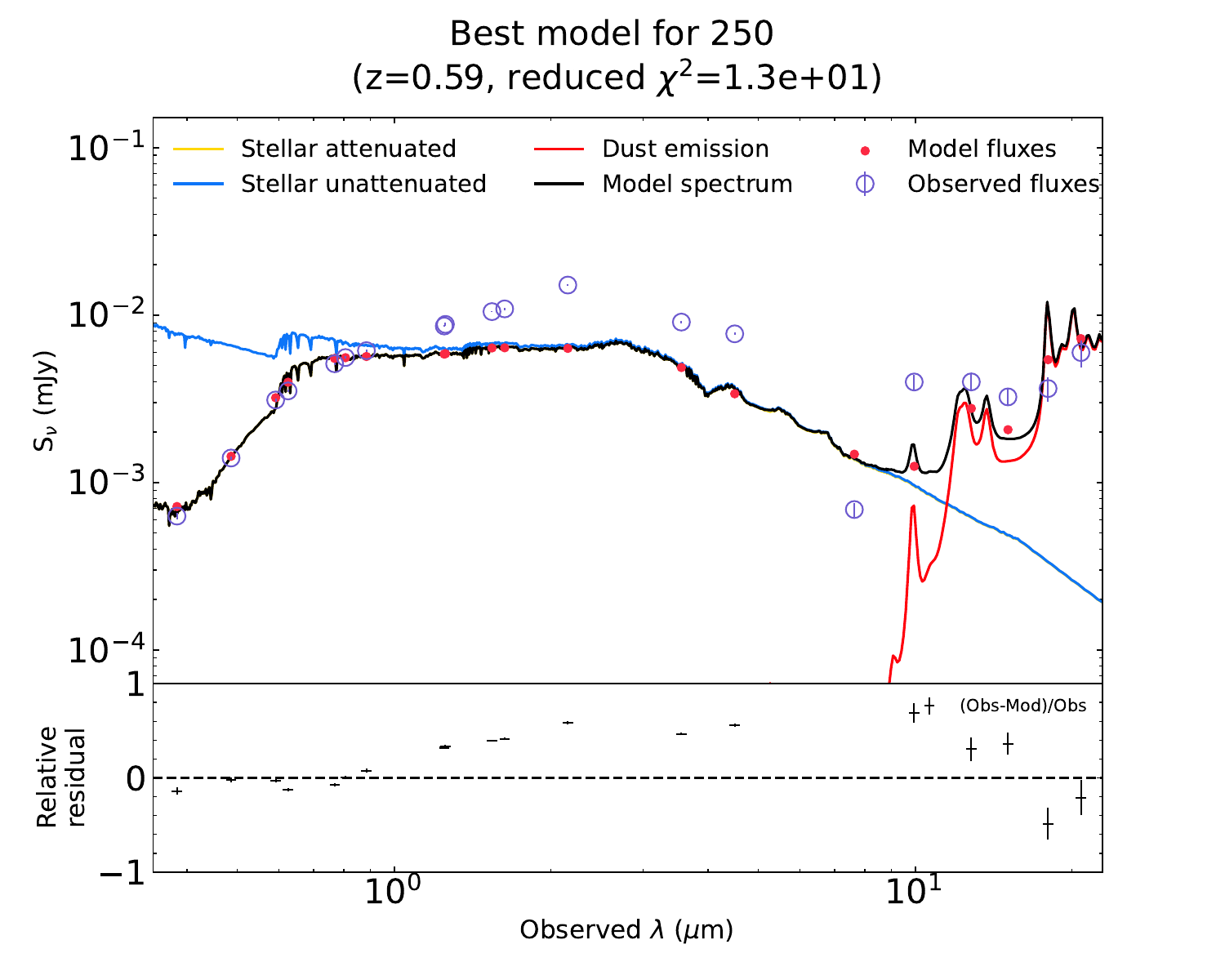}
    \caption{The same as Fig. \ref{fig:SED example1}, but for ID 250.}
    \label{fig:SED example12}
\end{figure}

In Sec.\ref{sec:results}, we present the SED fitting result in Fig. \ref{fig:SED example1}. We present other SED fit results in \ref{fig:SED example2}-\ref{fig:SED example12}. In Table \ref{tab:photometric_data}, we present six photometric data from \textit{JWST} MIRI observation.


\begin{table}
    
\centering
\caption{Flux densities and errors for PAH selected galaxies observed by \textit{JWST}. Note that the unit of flux densities is $\mu$Jy. There are 10 good fitting sources ($\chi<6$) and 2 bad fitting sources. The other 15 broad-band photometric data are from CANDELS-EGS catalog \citep{Finkelstein2017}, described in Sec.\ref{sec:data}.}
\label{tab:photometric_data}
\rotatebox{90}{
\begin{tabular}{rrrrrrrrrrrrrr}
\toprule
ID & Name in CANDELS\_EGS& \multicolumn{2}{c}{F770W} & \multicolumn{2}{c}{F1000W} & \multicolumn{2}{c}{F1280W} & \multicolumn{2}{c}{F1500W} & \multicolumn{2}{c}{F1800W} & \multicolumn{2}{c}{F2100W} \\
 & & $f_{\nu}$ & $\Delta f_{\nu}$ & $f_{\nu}$ & $\Delta f_{\nu}$ & $f_{\nu}$ & $\Delta f_{\nu}$ & $f_{\nu}$ & $\Delta f_{\nu}$ & $f_{\nu}$ & $\Delta f_{\nu}$ & $f_{\nu}$ & $\Delta f_{\nu}$ \\
\midrule

59 & CANDELS\_EGS\_F160W\_J142040.3+530306.6 & 3.45 & 0.30 & 3.10 & 0.33 & 23.65 & 2.88 & 37.55 & 3.92 & 31.14 & 4.39 & 14.41 & 1.94 \\ 
64 & CANDELS\_EGS\_F160W\_J142037.9+530249.5 & 1.20 & 0.11 & 0.74 & 0.09 & 3.46 & 0.43 & 7.20 & 0.76 & 5.97 & 0.91 & 10.12 & 1.49 \\ 
66 & CANDELS\_EGS\_F160W\_J142038.3+530254.9 & 11.93 & 1.03 & 10.00 & 1.05 & 50.01 & 6.08 & 118.25 & 12.33 & 99.52 & 14.00 & 15.86 & 2.10 \\ 
72 & CANDELS\_EGS\_F160W\_J142038.5+530306.1 & 2.93 & 0.25 & 2.36 & 0.25 & 13.92 & 1.69 & 28.60 & 2.98 & 29.40 & 4.15 & 7.39 & 1.23 \\ 
113 & CANDELS\_EGS\_F160W\_J142039.1+530351.7 & 4.70 & 0.41 & 4.82 & 0.51 & 13.58 & 1.65 & 36.80 & 3.84 & 211.84 & 29.79 & 57.72 & 7.08 \\ 
136 & CANDELS\_EGS\_F160W\_J142034.8+530329.6 & 7.93 & 0.69 & 10.09 & 1.06 & 28.72 & 3.49 & 78.93 & 8.23 & 146.53 & 20.61 & 33.05 & 4.11 \\ 
178 & CANDELS\_EGS\_F160W\_J142022.8+525859.5 & 8.14 & 0.70 & 6.70 & 0.70 & 20.89 & 2.54 & 44.31 & 4.62 & 94.34 & 13.27 & 28.74 & 3.60 \\ 
187 & CANDELS\_EGS\_F160W\_J142018.2+525825.4 & 7.38 & 0.64 & 6.38 & 0.67 & 18.66 & 2.27 & 48.55 & 5.06 & 108.60 & 15.27 & 30.71 & 3.84 \\ 
218 & CANDELS\_EGS\_F160W\_J142020.3+525925.1 & 2.49 & 0.22 & 1.67 & 0.18 & 10.17 & 1.24 & 19.46 & 2.03 & 16.74 & 2.38 & 3.56 & 0.96 \\ 
225 & CANDELS\_EGS\_F160W\_J142016.5+525859.9 & 1.02 & 0.09 & 0.87 & 0.10 & 4.50 & 0.56 & 10.22 & 1.08 & 7.00 & 1.04 & 14.32 & 1.93 \\ 

\bottomrule
145 & CANDELS\_EGS\_F160W\_J142033.6+530323.4 & 1.23 & 0.11 & 1.20 & 0.13 & 7.49 & 0.91 & 9.57 & 1.01 & 3.88 & 0.64 & 3.68 & 0.95 \\ 
250 & CANDELS\_EGS\_F160W\_J142017.0+525938.8 & 0.69 & 0.06 & 3.98 & 0.42 & 3.99 & 0.49 & 3.24 & 0.37 & 3.64 & 0.61 & 5.97 & 1.10 \\

\bottomrule
\end{tabular}
}
\end{table}


\bsp	
\label{lastpage}
\end{document}